\documentclass[aps,floats,prd,twocolumn,nofootinbib,superscriptaddress]{revtex4-2}

\usepackage{amssymb}
\usepackage{amsmath}

\usepackage{graphicx}
\usepackage{graphics}
\usepackage{color}
\usepackage{fancyhdr} 
\usepackage{graphicx}
\usepackage{float}

\usepackage{subfig}
\usepackage[colorlinks=true,linktocpage=true,linkcolor=blue,citecolor=blue]{hyperref}

\usepackage[utf8]{inputenc}

\usepackage[justification=centerlast, format=plain, labelfont=bf]{caption}

    \newcommand{\be}{\begin{equation}}
  \newcommand{\ee}{\end{equation}}
    \newcommand{\ba}{\begin{align}}
  \newcommand{\ea}{\end{align}}

\newcommand{\Msun}{M_{\odot}}
\newcommand{\Mpcinv}{ {\rm Mpc}^{-1} }

\newcommand{\mnras}{Mon. Not. R. Astron. Soc.}

\newcommand{\jcap}{J. Cosmol. Astropart. Phys.}

\begin{document}

\title{ETHOS - An Effective Theory of Structure Formation:\\  Impact of Dark Acoustic Oscillations on Cosmic Dawn}

\author{Julian B.~Mu\~noz}
\affiliation{Harvard-Smithsonian Center for Astrophysics, 60 Garden St., Cambridge, MA 02138, USA}
\affiliation{Department of Physics, Harvard University, 17 Oxford St., Cambridge, MA 02138, USA}
\author{Sebastian Bohr}
\affiliation{Centre for Astrophysics and Cosmology, Science Institute, University of Iceland, Dunhagi 5, 107 Reykjavik, Iceland}
\author{Francis-Yan Cyr-Racine}
\affiliation{Department of Physics and Astronomy, University of New Mexico, Albuquerque, NM 87106, USA}
\author{Jes\'us Zavala}
\affiliation{Centre for Astrophysics and Cosmology, Science Institute, University of Iceland, Dunhagi 5, 107 Reykjavik, Iceland}
\author{Mark Vogelsberger}
\affiliation{Department of Physics, Kavli Institute for Astrophysics and Space Research, Massachusetts Institute of Technology, Cambridge, MA 02139, USA}

\begin{abstract}
Upcoming data of the 21-cm hydrogen line during cosmic dawn ($z\sim 10-30$) will revolutionize our understanding of the astrophysics of the first galaxies.
Here we present a case study on how to exploit those same measurements to learn about the nature of dark matter (DM) at small scales.
Focusing on the Effective Theory of Structure Formation (ETHOS) paradigm, we run a suite of simulations covering a broad range of DM microphysics, connecting the output of $N$-body simulations to dedicated 21-cm simulations to predict the evolution of the 21-cm signal across the entire cosmic dawn.
We find that observatories targeting both the global signal and the 21-cm power spectrum are sensitive to all ETHOS models we study, and can distinguish them from CDM if the suppression wavenumber is smaller than $k\approx 300\, h/$Mpc, even when accounting for feedback with a phenomenological model. 
This is an order of magnitude smaller comoving scales than currently constrained by other data sets, including the Lyman-$\alpha$ forest.
Moreover, if a prospective 21-cm detection confirmed a deficiency of power at small scales, we show that ETHOS models with strong dark acoustic oscillations can be discriminated from the pure suppression of warm dark matter, showing the power of 21-cm data to understand the behavior of DM at the smallest physical scales.
\end{abstract}

\maketitle

\section{Introduction}

The majority of matter in our universe is dark, and seemingly collisionless~\cite{1981ApJ...250..423D,Blumenthal:1982mv,Blumenthal:1984bp,Davis:1985rj,1984ApJ...277..470P,Bertone:2004pz}.
Decades of observational efforts have provided us with increasingly precise constraints on the nature of dark matter (DM)~\cite{Abbott:2017wau,Aghanim:2018eyx,2018PhRvD..98h3540M,Nadler:2019zrb,Nadler2020,Vogelsberger:2019ynw}, albeit not a solution to its nature yet.
An exciting possibility is that a complex dark sector hosts dark matter, as well as other components, which may interact with each other throughout cosmic history~\cite{Foot:2004pa,Ackerman:2008gi, ArkaniHamed:2008qn,Feng:2009mn,Kaplan:2009de,Behbahani:2010xa,Kaplan:2011yj,Aarssen:2012fx,Cline:2012is,Hooper:2012cw,Das:2012aa,Cyr-Racine:2013ab,Diamanti:2012tg,Baldi:2012ua,Fan:2013yva,Fan:2013tia,McCullough:2013jma,Cline:2013pca,Cline:2013zca,Bringmann:2013vra,Chu:2014lja,Archidiacono:2014nda,Randall:2014kta,Buen-Abad:2015ova,Lesgourgues:2015wza,Choquette:2015mca}.

Searching for these dark-sector interactions between DM and light degrees of freedom, while impossible in the lab, is feasible with astrophysical data sets (see e.g.~Ref.~\cite{Tulin:2017ara}).
DM interactions can leave an imprint on the formation of cosmic structure, which can be searched with precision cosmic data sets such as the cosmic microwave background (CMB) and large-scale structure (LSS) of the universe ~\cite{Cyr-Racine:2013fsa,Archidiacono:2017slj,Archidiacono:2019wdp}. Past analyses have shown these cosmological data sets to be broadly consistent with the standard cold dark matter (CDM) paradigm on large scales. Any significant departure from the ``vanilla" CDM behavior thus ought to appear preferentially at smaller scales.
In this regime, observations of the Lyman-$\alpha$ forest \cite{2018PhRvD..98h3540M}, of the luminosity function of Milky Way satellites \cite{Nadler:2019zrb,Nadler2020}, and of flux-ratio anomalies of multiply imaged strongly lensed quasars \cite{Dalal:2001fq,Gilman:2019vca,Hsueh:2019ynk,Gilman:2019nap} have shown consistency with CDM on halo mass scale $\gtrsim 10^9\Msun$. Pushing this boundary to even smaller scales is a major goal of a current and future multi-pronged effort (see e.g.~Ref.~\cite{Drlica-Wagner:2019xan}).

A telltale signature of DM interacting with light degrees of freedom in the early Universe is the presence of dark acoustic oscillations (DAOs) in the linear matter power spectrum. Detailed simulations \cite{Buckley:2014hja,Schewtschenko:2014fca,2016MNRAS.460.1399V} of the nonlinear evolution of structure within such models have shown that this key signature gets partially erased as power is regenerated on small scales at late times. Therefore, observations at higher redshifts have the potential to probe DAOs and their effect on structure formation in a more pristine state. One of the earliest probe of nonlinear structure formation in our Universe is the 21-cm signal from cosmic dawn. At that epoch, the ultraviolet (UV) radiation emitted by the first stars recouples the neutral hydrogen spin temperature to that of the cooler gas via the Wouthuysen-Field effect \cite{Wout,Field,Hirata:2005mz}, leading to a net absorption of 21-cm photons from the Rayleigh-Jeans tail of the CMB. Since early stellar formation depends sensitively on the abundance and properties of small DM halos with mass $M_{\rm h}\sim 10^6-10^8\Msun$, the timing and shape of this absorption feature can be used to search for the presence of DAOs and related damping on those scales.

In general, any model which suppresses or modifies the amplitude of DM fluctuations on small scales could affect the 21-cm cosmic dawn signal (see e.g.~Refs.~\cite{Lopez-Honorez:2016sur,Schneider:2018xba,Lopez-Honorez:2018ipk,Escudero:2018thh,Munoz:2018jwq,Munoz:2018pzp,Yoshiura:2019zxq,Mena:2019nhm,Munoz:2019hjh}). Exploring the 21-cm signal from this broad parameter space of possible DM models can be quite costly since it generally requires detailed simulations. A promising approach is to map the different DM microphysics to effective parameters that govern how structure forms. The effective theory of structure formation (ETHOS) \cite{Cyr-Racine:2015ihg,Vogelsberger:2015gpr} provides such a mapping. It naturally interpolates between DM models having sharp transfer function cutoff such as warm DM (WDM) to theories displaying damped or strong acoustic oscillations, and to models looking nearly like CDM. So far, the ETHOS framework has been used to study the satellite galaxies of Milky Way-like hosts \cite{Vogelsberger:2015gpr}, the high-redshift UV luminosity function and reionization \cite{Lovell:2017eec}, and the impact of DAOs on Lyman-$\alpha$ forest signal \cite{Bose:2018juc}.

In this paper, we use the simple but powerful phenomenological ETHOS parametrization introduced in Ref.~\cite{Bohr:2020yoe} to describe deviations from the standard CDM scenario and compute the expected 21-cm signal from cosmic dawn. This two-dimensional parameter space spans a broad range of models ranging from WDM and models with suppressed DAOs, to models displaying strong DAOs and theories that are phenomenologically undistinguishable from CDM. Using this parametrization, we compute both the expected 21-cm global signal and power spectrum and study the distinguishability of different dark matter models in upcoming experiments.

This paper is structured as follows. In Sec.~\ref{sec:ETHOS} we describe the ETHOS parametrization and the $N$-body simulations we use.
We show the effect of the different ETHOS models on the 21-cm global signal in Secs.~\ref{sec:21cmGS}, and on the 21-cm fluctuations in ~\ref{sec:21cmPS}.
We conclude in Sec.~\ref{sec:Conclusions}.

\section{The ETHOS framework and simulations}
\label{sec:ETHOS}

Here we describe the matter power spectrum  within the ETHOS framework, and the simulations that we use.

\subsection{Effective parametrization}

The ETHOS paradigm was developed to capture the effects of DM microphysics on the formation of structure in our universe in a few convenient parameters~\cite{Cyr-Racine:2015ihg}. Throughout this work we will employ the effective ETHOS parametrization introduced in Ref.~\cite{Bohr:2020yoe}, which provides a convenient---and accurate---shortcut to the full ETHOS parameter space.

\begin{figure}[t!]
	\includegraphics[width=0.44\textwidth]{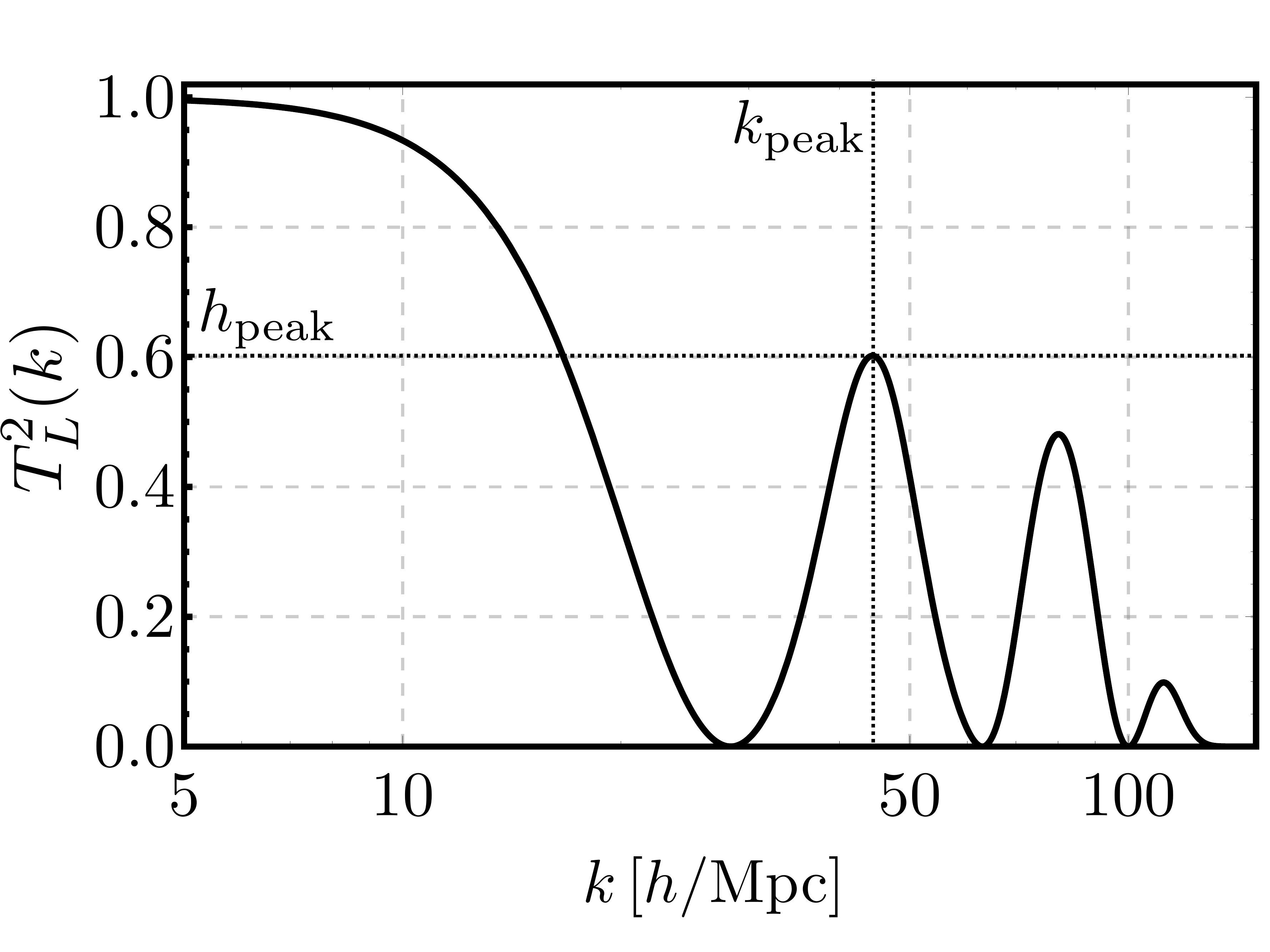}
	\caption{Diagram of the ETHOS parametrization of the power spectrum.
	Shown is the linear ``transfer" function $T_L^2=P_{m}^{\rm ETHOS}/P_{m}^{\rm CDM}$ as a function of wavenumber $k$.
	The two parameters determine the location $k_{\rm peak}$ and height $h_{\rm peak}$ of the first peak, where $h_{\rm peak}=0$ corresponds to WDM with different masses, and $k_{\rm peak}\to\infty$ to CDM.
	}	
	\label{fig:diagram}
\end{figure}

This circumvents modeling the DM interactions, and instead approximates the matter power spectrum through two relevant parameters, which control the height $h_{\rm peak}$ and wavenumber $k_{\rm peak}$ of the first DAO peak, as illustrated in Fig.~\ref{fig:diagram}.
In this notation the limit $h_{\rm peak}\to 0$ corresponds to WDM, whereas $h_{\rm peak}\to 1$ are strong DAOs.
As an example, an atomic-DM model will have $h_{\rm peak} \to 0$ if diffusion damping occurs at larger scales than the DAOs, and $h_{\rm peak} \to 1$ if dark recombination occurs instantaneously. 
These two parameters capture the main features of the matter power spectrum for a large variety of ETHOS models (which include more details about the DM microphysics), and  it was shown in Ref.~\cite{Bohr:2020yoe} that the high-redshift halo mass function (HMF) is well approximated with only these two degrees of freedom.

The connection between these phenomenological parameters and particle physics model parameters (masses, couplings, etc.) is provided in Ref.~\cite{Bohr:2020yoe}.
For instance, the $h_{\rm peak}=0$ cases are equivalent to a WDM mass
\be
\frac{m_{\rm WDM}}{1\,{\rm keV}} = \left[0.050 \left(\frac{k_{\rm peak}}{h\,{\rm Mpc}^{-1}}\right) \left(\frac{\Omega_\chi}{0.25}\right)^{0.11} \left(\frac{h}{0.7}\right)^{1.22} \right]^\frac{1}{1.11},
\label{eq:kpeakWDM}
\ee
where $\Omega_\chi$ is the DM abundance.
We use the same models as Ref.~\cite{Bohr:2020yoe} in this work, i.e., 48 simulations with $h_{\rm peak}=0-1$ in steps of 0.2 and $k_{\rm peak}=35-300$\,$h/$Mpc (where $h$ is the reduced Hubble constant) with equidistant steps in $\log(k_{\rm peak})$ on the intervals [35,100]\,$h/$Mpc and [100,300]\,$h/$Mpc.

\subsection{$N$-body Simulations}

We run cosmological DM-only $N$-body simulations with the code \textsc{Arepo}~\cite{Springel2010} using the zoom-in technique described in Ref.~\cite{Bohr:2020yoe} with a particle mass of $8\times10^4\,{\rm M_\odot}h^{-1}$ in the high-resolution region. The initial conditions are generated by the code \textsc{MUSIC}~\cite{Hahn2011} and the cosmological parameters of the simulations are $\Omega_{\rm m}=0.31069$, $\Omega_\Lambda=0.68931$, $H_0=67.5\,{\rm km/s/Mpc}$, $n_{\rm s} = 0.9653$ and $\sigma_8=0.815$, where $\Omega_{\rm m}$ and $\Omega_\Lambda$ are the fraction of the matter-energy density of the Universe today, that is provided by matter and cosmological constant, respectively, $H_0$ is the Hubble constant today, $n_{\rm s}$ is the spectral index, and $\sigma_8$ is the mass variance of linear fluctuations in 8\,$h^{-1}$\,Mpc spheres at $z=0$.

The output we will use are the HMFs measured at each redshift in the range $z=10-25$ with redshift intervals $\Delta z=0.3$, which are passed as an input to our modified version of {\tt 21cmvFAST}, as we will describe below.
We find the HMF through counting the number of haloes identified by the friends-of-friends and Subfind algorithm in \textsc{Arepo} within the high-resolution region of the simulation.

\subsection{Ingredients for the 21-cm Simulations}

Let us now describe how we use the ETHOS results from above to find the evolution of the 21-cm signal across cosmic dawn.
In this work we will use semi-numerical 21-cm simulations with a modified version of the public code {\tt 21cmvFAST}~\cite{Munoz:2019rhi,Munoz:2019fkt}\footnote{\url{https://github.com/JulianBMunoz/21cmvFAST}}, which itself is based on {\tt 21cmFAST}~\cite{Mesinger:2010ne,Greig:2015qca}\footnote{\url{https://github.com/andreimesinger/21cmFAST}}.
Here, however, we do not assume the HMF of a CDM model.
Instead, we use the HMF from the ETHOS simulations, denoted as $dn/dM$, to compute the fraction of baryons collapsed into stars as
\be
F_{\rm coll} = \int_{\rm M_{\rm cool}}^\infty dM \dfrac{M}{\rho_m} \dfrac{dn}{dM} \dfrac{f_g}{f_b} f_*(M),
\label{eq:fcoll}
\ee
$f_b$ and $f_g$ are the baryon and gas fractions, and $f_*$ is the fraction of gas that gets converted onto stars.
This integral runs over masses larger than $M_{\rm cool}$, which parametrizes the smallest halo that can form stars efficiently (note that an alternate parametrization exponentially suppresses low-mass haloes, instead of providing a sharp cut-off, providing similar results~\cite{Park:2018ljd}).
Throughout this work we assume, for simplicity, that only haloes above the atomic-cooling threshold can form stars, i.e., $M_{\rm cool}=M_{\rm atom}(z)$~\cite{Oh:2001ex}.
This provides a conservative estimate of the reach of cosmic-dawn data to probe
ETHOS models, as smaller (molecular-cooling) haloes would be further affected by deviations from CDM.

\begin{figure}[btp!]
	\includegraphics[width=0.5\textwidth]{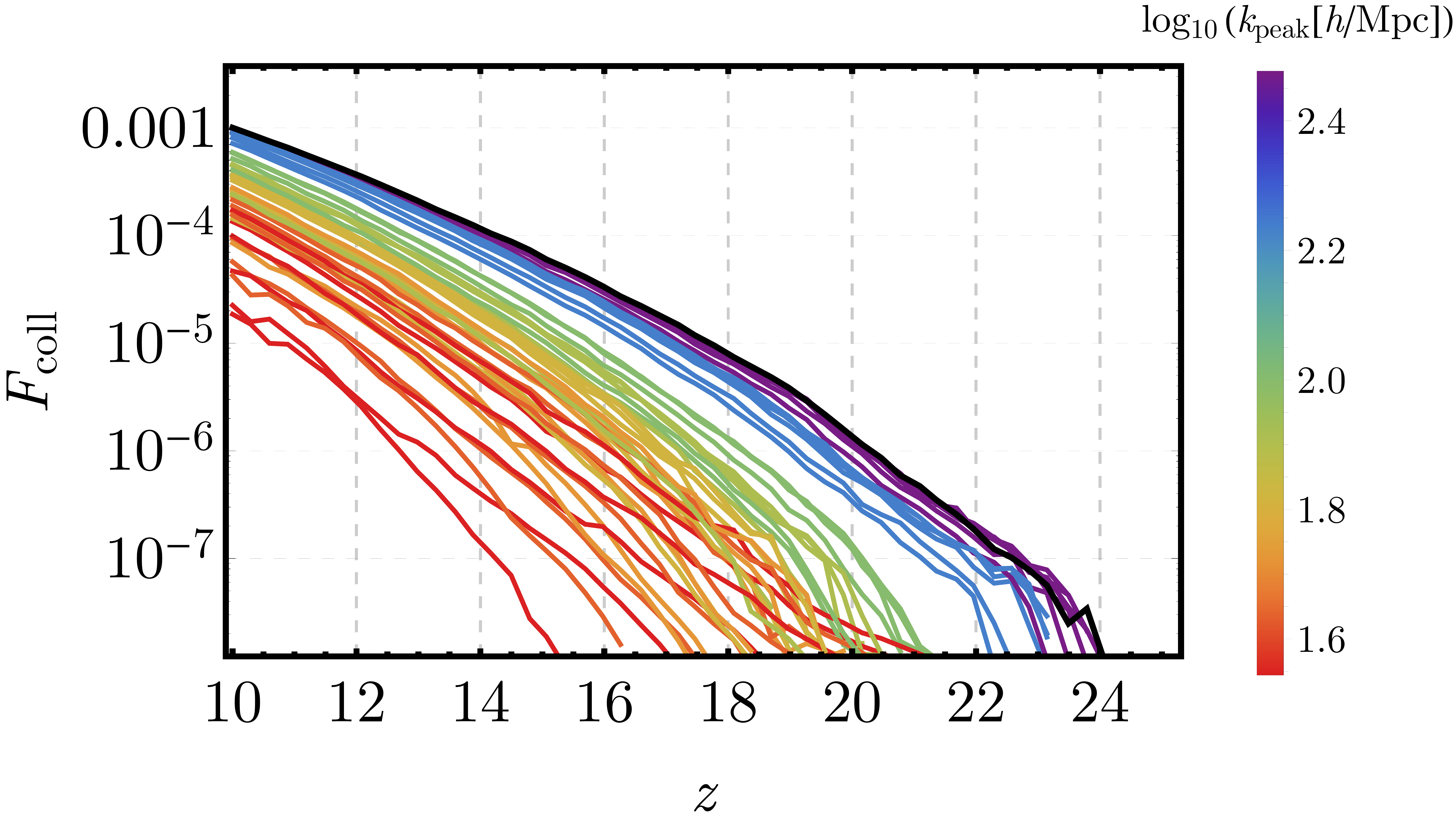}
	\caption{Collapsed fraction of baryons to star-forming haloes as a function of redshift $z$, for all our simulations.
	In all cases we assume that haloes above the atomic-cooling threshold can form stars, and consider no further feedback in this plot.
	Lines are colored by the wavenumber of their first peak $k_{\rm peak}$, regardless of the height $h_{\rm peak}$, with CDM corresponding to the highest $k_{\rm peak}$ shown.
	The black line corresponds to CDM.
	}	
	\label{fig:Fcoll}
\end{figure}

In practice we evaluate Eq.~\eqref{eq:fcoll} by directly adding the mass of haloes above $M_{\rm cool}(z)$, to avoid errors induced by binning the HMF.
We show the resulting $F_{\rm coll}$ as a function of redshift for all our ETHOS models, and CDM, in Fig.~\ref{fig:Fcoll}.
As expected, this quantity grows exponentially for all models as the cosmic evolution makes fluctuations grow bigger, and more haloes form.
However, models with low $k_{\rm peak}$ take significantly longer to form galaxies, shifting all their lines to lower $z$.
We note, in passing, that for very low values of $F_{\rm coll}$ (corresponding to high redshifts) the Poisson noise is important for all models.
This causes the $F_{\rm coll}$ curves of some ETHOS models to overcome the CDM case, albeit only briefly and at very high $z$.

As we neglect molecular-cooling haloes, the main source of feedback to consider is photo-heating, which can evaporate the gas within haloes~\cite{Efstathiou:1992zz,Dijkstra:2003vg}.
However, atomic-cooling haloes are not expected to be significantly affected by photo-heating feedback until $z\sim 10$~\cite{Sobacchi:2013ww,Qin:2020xyh,Qin:2020pdx}, where we stop our simulations.
To account for any residual feedback (such as due to SNe), we will implement a model where the gas fraction that turns into stars as~\cite{Trenti:2010sz, Sitwell:2013fpa, Mason:2015cna,Park:2018ljd}.
\be
f_*(M) = f_*^{(0)} \times  \left(\dfrac{M}{M_0}\right)^\alpha,
\label{eq:fstaralpha}
\ee
where we take $f_*^{(0)}=0.1$ at a scale $M_0=1.6\times10^{11}\,\Msun$ (note that this power-law behavior is expected to break for higher-masses haloes~\cite{Tacchella:2018qny,Trenti:2010sz}, which however do not significantly affect the 21-cm signal during cosmic dawn). 
While this simplistic model is not expected to capture all the complexities of feedback in the first galaxies, it will serve to study the impact of feedback on our models.

We will conservatively assume that $\alpha=0$ for all ETHOS models, as further feedback would only make them deviate more from CDM.
For CDM, on the other hand, we will vary $\alpha$ in the range $[0-0.5]$, in order to estimate the impact of feedback, and whether the different ETHOS models can be distinguished from it.
We note that our range of values of $\alpha$ is lower than typical of lower-$z$ probes, such as galaxy luminosity functions, where $\alpha\approx 1$~\cite{Gillet:2019fjd,Tacchella:2018qny,Yung_2018}, as we expect feedback to be less important during cosmic dawn.

As our ETHOS HMFs are obtained exclusively from a zoom-in region within a larger simulation box (see Ref.~\cite{Bohr:2020yoe}), we need to apply a correction for the possible difference in mean density between the zoom region and the whole cosmological volume. To do so, we use an extended Press-Schechter formalism \cite{Press:1973iz} in which we rescale the collapsed fractions as
\be
F_{\rm coll} (z) \to F_{\rm coll} (z) \dfrac{{\rm erfc}\left[\dfrac{\delta_{\rm crit} - \delta_{\rm zoom}(z)}{\sqrt{2 S(z)}}\right]  } {{\rm erfc}\left[\dfrac{\delta_{\rm crit} }{\sqrt{2 S(z)}}\right] }
\ee
given the overdensity $\delta_{\rm zoom}$ in the zoom-in region (measured in the simulations), where $\delta_{\rm crit} = 1.686$ is the critical density for collapse, and
\be
S (z) = \sigma_{\rm cool}^2(z) - \sigma_{\rm zoom}^2(z)
\ee
is the variance on the cooling haloes, corrected by that in patches of the zoom-in region, $\sigma^2_{\rm zoom}$.
We expand on how we compute the variances below.
We have confirmed that this procedure reproduced the collapsed fraction in zoom-in simulations with average density for CDM.
Moreover, the standard procedure used in {\tt 21cmFAST} is to modify $F_{\rm coll}$ in over/under-dense pixels via this same formula, so our re-scaling would be equivalent to changing the average density of the overall {\tt 21cmFAST} box to be $\delta_{\rm zoom}(z)$ and using the zoom-in overdense $F_{\rm coll}$.

The other ingredient modified in ETHOS models is the matter power spectrum, which changes the variance $\sigma^2$ of fluctuations on different scales.
Since all ETHOS models we consider follow CDM at large scales the variance on the pixel size is not altered.
Nevertheless, the variance on the scale at which atomic-cooling haloes form will change.
We calculate it as
\be
\sigma_{\rm cool}^2(z) = \int \dfrac{d^3 k}{(2\pi)^3} P_m(k) |\mathcal W(k R_{\rm cool})|^2,
\ee
where $R_{\rm cool}=R_{\rm atom}(z)$ is the comoving radius of atomic-cooling haloes at each $z$, and $\mathcal W$ is a window function, which can have different functional forms, such as a (real-space) top-hat.
Nevertheless, it has been shown that the HMFs of non-CDM models are better fit when using a smooth window function
\be
\mathcal W(x) = \dfrac{1}{1+(x/c)^\beta},
\ee
with $c=3.7$ and $\beta=3.5$,
as calibrated in Ref.~\cite{Sameie:2018juk} to fit the HMF of models with DAO, such as the ones we study here.

We note that we conservatively do not alter the reionization calculation from {\tt 21cmFAST}, as we are interested in the cosmic-dawn era only.
We encourage the reader to see Refs.~\cite{Lovell:2017eec,Lovell:2018gap,Bose:2018juc} for the effect of ETHOS models on reionization and the Lyman-$\alpha$ forest.

\section{Effect on the 21-cm Global Signal}
\label{sec:21cmGS}

The different histories of early structure formation of each of the ETHOS models will give rise to different 21-cm signals during cosmic dawn.
Here we explore this observable, starting with the global signal---the average absorption or emission of 21-cm photons across the entire sky at each frequency or redshift.
This signal has been targeted by several experiments~\cite{LEDA,Singh:2017syr,PRIZM,Voytek:2013nua,DiLullo:2020owx}, including a first detection claimed by the EDGES collaboration~\cite{Bowman:2018yin}.

\subsection{The observable}

We define the usual 21-cm brightness temperature as,
\ba
T_{21}(\mathbf x) =& \, 38\,{\rm mK}
\left(1 - \dfrac{T_\gamma}{T_S}\right)
\left(\dfrac{1+z}{20}\right)^{1/2} \nonumber \\
&\times \left(\dfrac{\partial_r v_r}{H}\right)^{-1} 
x_{\rm HI} (1+\delta_b) 
,
\end{align}
where $\partial_r v_r$ is the radial velocity gradient, $H(z)$ is the Hubble expansion rate, $\delta_b$ is the baryonic overdensity, and $T_\gamma$ and $T_S$ are the photon (CMB) and spin temperatures, respectively.
During the cosmic-dawn era that we are interested in the hydrogen neutral fraction $x_{\rm HI}\approx 1$.
For a thorough review of the physics of the 21-cm line  we refer the reader to Refs.~\cite{Pritchard:2011xb,Furlanetto:2006jb}.
The 21-cm temperature will be computed at each point using the {\tt 21cmvFAST} simulations outlined above, and the global signal $\overline{T_{21}}$ is obtained by simply averaging the entire box output at each redshift.

Throughout this work we will use a single set of initial conditions for all the simulations, to ease comparison, generated with the {\it Planck} 2018 best-fit cosmological parameters~\cite{Aghanim:2018eyx}.
Moreover, we will fix the astrophysical parameters to be the same as in Ref.~\cite{Munoz:2019hjh}.
Our simulation boxes have 600 Mpc comoving in size, and 3 Mpc resolution, and are ran from $z=35$ to $z=10$, to avoid the bulk of reionization.

\begin{figure}[t!]
	\includegraphics[width=0.5\textwidth]{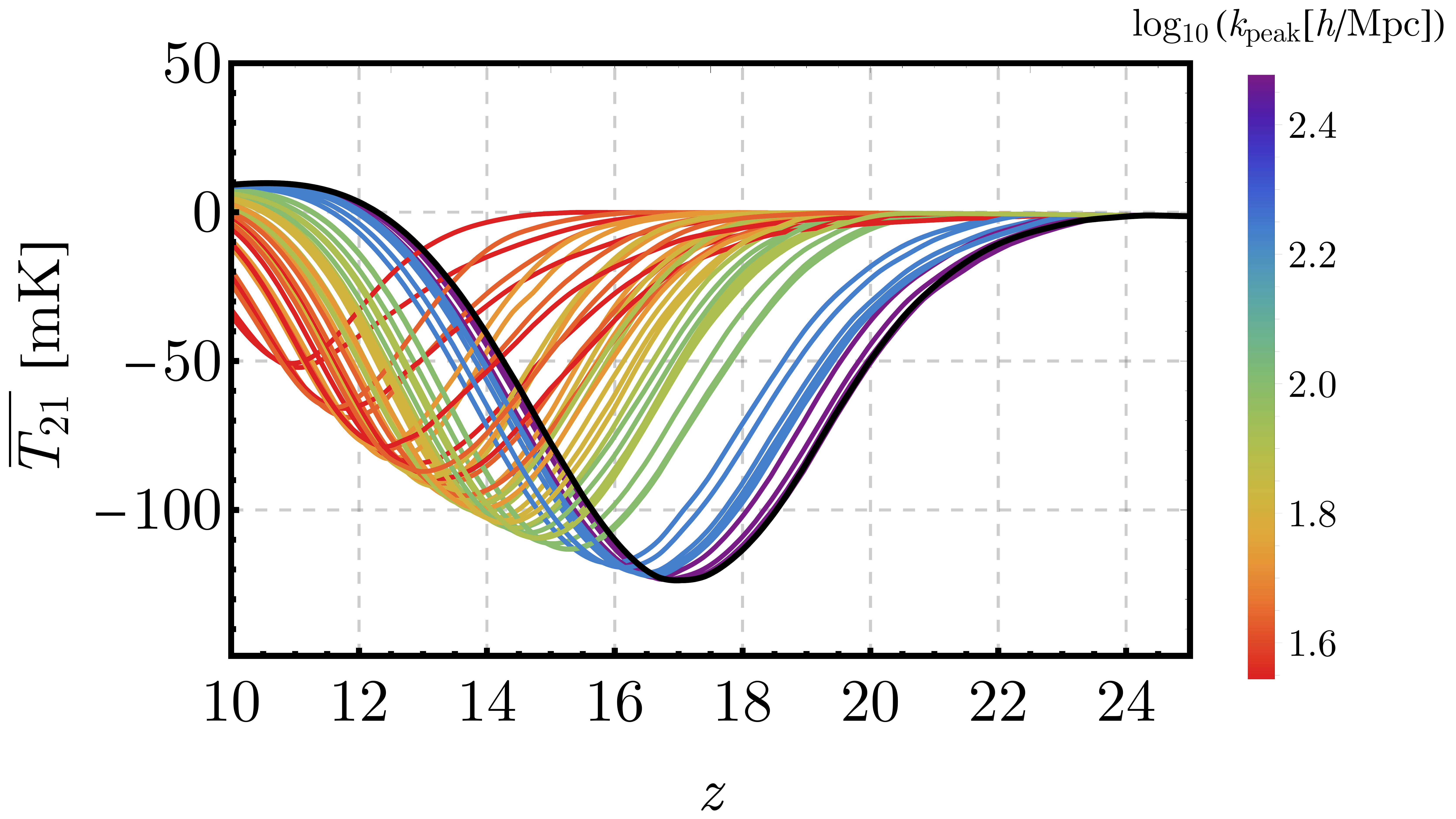}
	\caption{Global signal as a function of redshift for all our ETHOS simulations.
	As in Fig.~\ref{fig:Fcoll}, the color scale indicates the scale $k_{\rm peak}$ of the first peak, and black corresponds to CDM.
    All models show the same landmarks of evolution, explained in the main text, although the location and depth of the peaks change between models. 
	}	
	\label{fig:GS_allETHOS}
\end{figure}

Under these conditions, we show the output of all of our ETHOS models, and CDM, in Fig.~\ref{fig:GS_allETHOS}.
Their overall evolution can be summarized as follows. 
First, during the onset of the Lyman-$\alpha$ coupling era (LCE; at $z\sim 22$ for CDM) the GS deviates from zero due to the UV photons from the first stars, which produce Wouthuysen-Field coupling~\cite{Wout,Field,Hirata:2005mz}.
Second, the transition from the LCE to the epoch of heating (EoH; at $z\sim 17$ for CDM) the signal starts growing due to the X-ray heating of the neutral hydrogen~\cite{Pritchard:2006sq,Pacucci:2014wwa}.
Finally, the EoH gives way to the epoch of reionization (EoR; at $z\sim 12$ for CDM) where the IGM is fully heated and the signal is reduced as hydrogen slowly becomes ionized~\cite{Barkana:2000fd,Pritchard:2008da}.

\begin{figure}[t!]
	\includegraphics[width=0.5\textwidth]{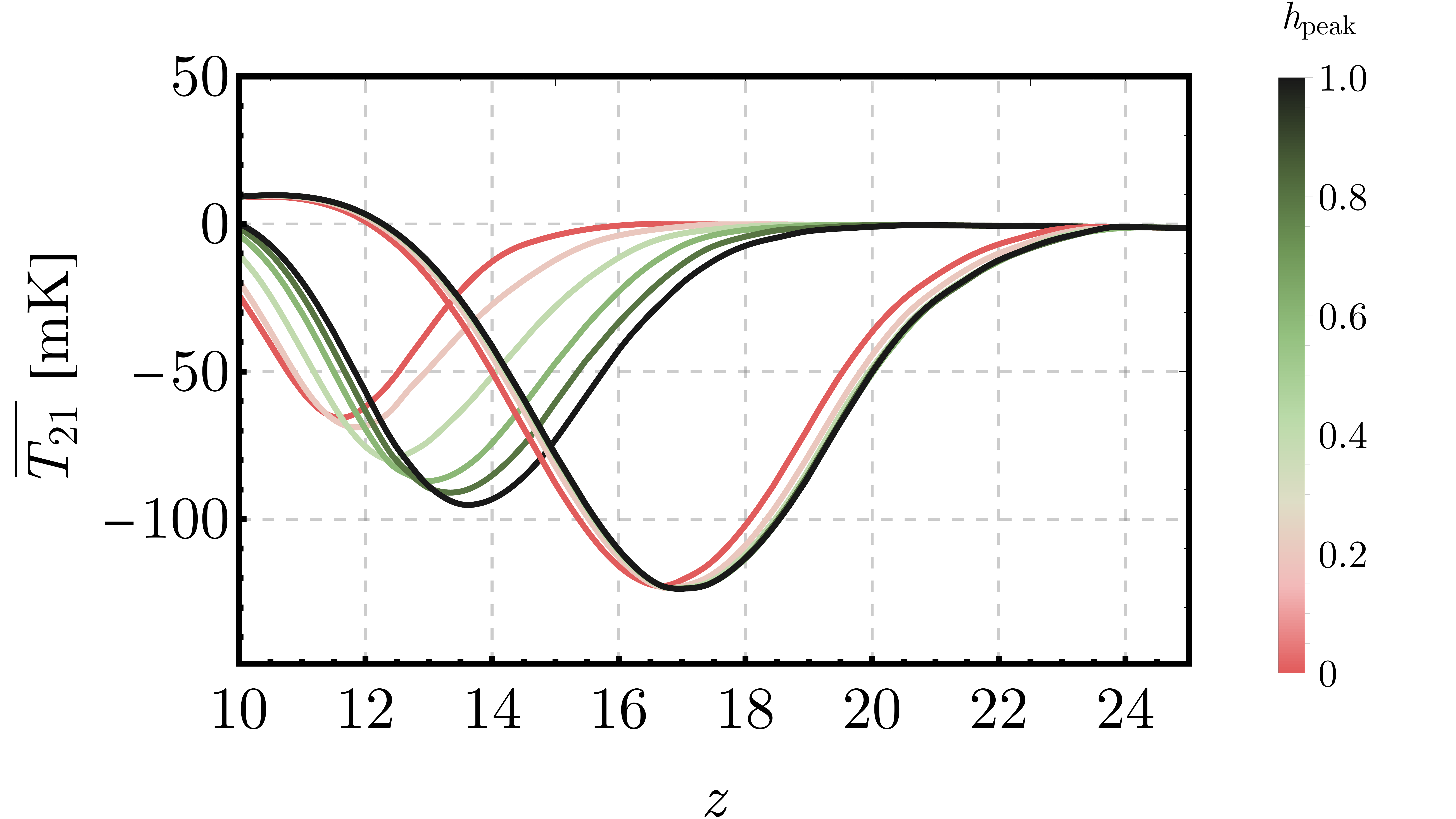}
	\caption{Same as Fig.~\ref{fig:GS_allETHOS} but for only models with $k_{\rm peak}= 43 \, h/$Mpc (left lines) and 300 $h/$Mpc (right lines), where the color indicates the value of $h_{\rm peak}$.
	Models closer to WDM ($h_{\rm peak}\sim 0$) have less structure formation, and thus a delayed 21-cm absorption signal.
	}	
	\label{fig:GS_hpeak}
\end{figure}

While all the models shown in Fig.~\ref{fig:GS_allETHOS} exhibit a similar overall evolution, models with more suppressed power are delayed with respect to CDM.
Furthermore, the entire shape of the GS depends on the details of the initial power spectrum, as models with additional power at large $k$ produce a more quickly evolving 21-cm global signal at high $z$.
To illustrate this point, we show in Fig.~\ref{fig:GS_hpeak} the GS for a subset of models with $k_{\rm peak}= 43$ and 300 $h/$Mpc, for different values of $h_{\rm peak}$.
Stronger DAOs (higher $h_{\rm peak}$) produce less suppression in the HMF, and thus an earlier 21-cm evolution.
This effect is more apparent for low $k_{\rm peak}$, as the haloes observed probe a broader range of $k$ in the matter power spectrum.
As we will explore below, this will allow us to distinguish different ETHOS models from one another.

\begin{figure}[hbtp!]
	\includegraphics[width=0.5\textwidth]{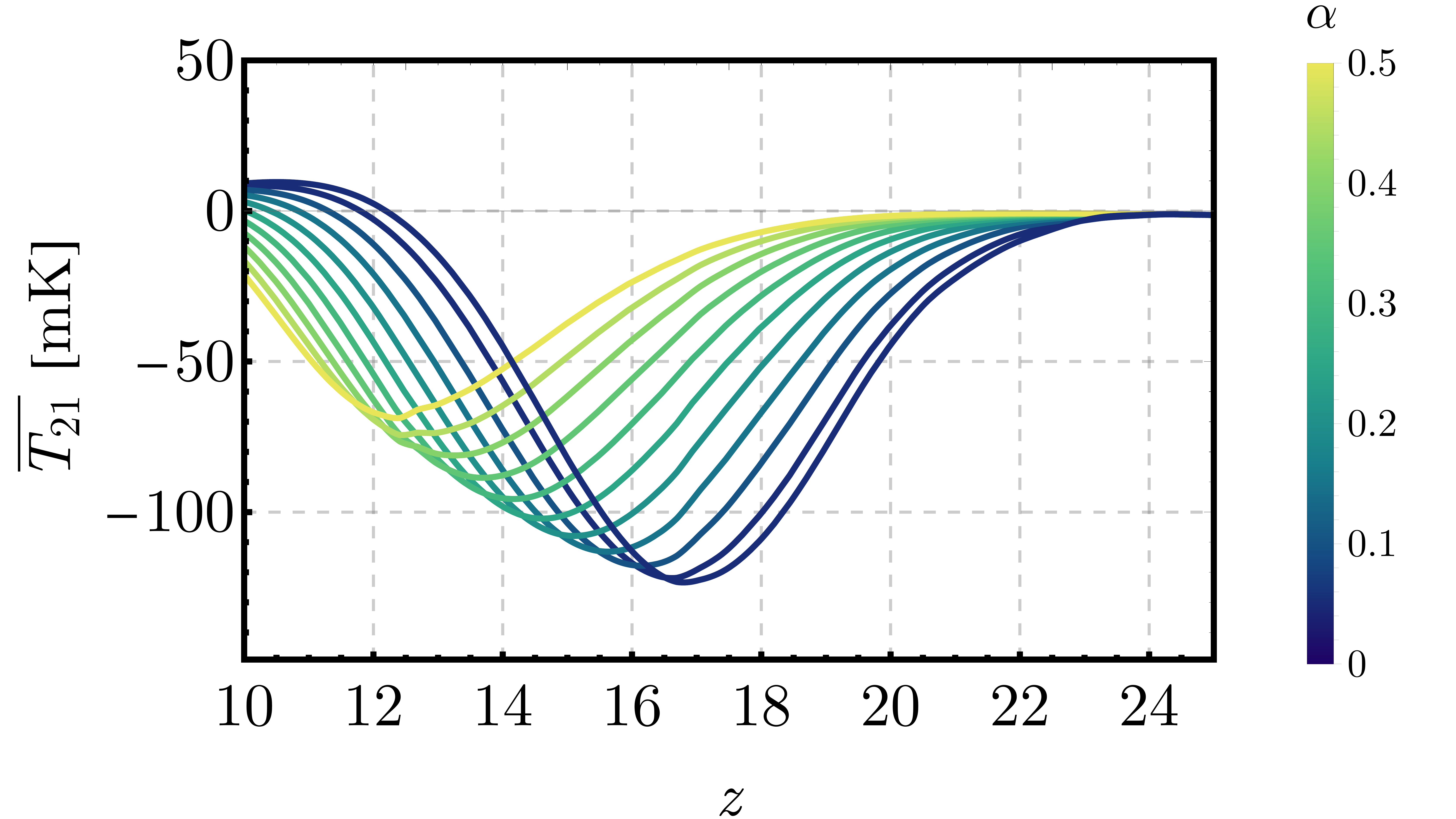}
	\caption{Same as Fig.~\ref{fig:GS_allETHOS} but for CDM only, where we vary the feedback parameter $\alpha$ from Eq.~\eqref{eq:fstaralpha}.
	Larger $\alpha$ corresponds to stronger feedback, and thus to a delayed 21-cm absorption signal.
	}	
	\label{fig:GS_alpha}
\end{figure}

Finally, we are also interested in distinguishing ETHOS models from CDM in the presence of feedback.
As described in the previous section, we will phenomenologically account for further possible sources of feedback by varying the parameter $\alpha$ in Eq.~\eqref{eq:fstaralpha}, which suppresses the amount of star formation in a mass-dependent way for each halo.
We show how the 21-cm GS varies with increasing $\alpha$ in Fig.~\ref{fig:GS_alpha}, which trivially delays the evolution of the GS.
Note that this delay is relatively smooth, as opposed to the sharper delay shown in Fig.~\ref{fig:GS_hpeak}, especially for $h_{\rm peak}=0$ (WDM) models.
This is to be expected, as this power-law-like astrophysical feedback does not cut off all haloes below some scale, whereas the ETHOS models approximately do.
This will help us to differentiate ETHOS models from CDM+feedback.

\subsection{Detectability}

While we have shown that different ETHOS models show very different 21-cm signals as a function of redshift, given their different amounts of structure formation, we have not shown whether this effect can be mimicked by feedback, and if different ETHOS models can be distinguished from each other (as for instance models with stronger DAOs and a lower $k_{\rm peak}$ can produce similar amounts of suppression as WDM with higher $k_{\rm peak}$, see Fig.~\ref{fig:GS_hpeak}).
We now perform a simple analysis to find how differentiable ETHOS models are from each other and from CDM, even when including potential feedback.

A realistic analysis should simultaneously fit for the cosmological 21-cm signal as well as the Galactic, extra-Galactic, and atmospheric foregrounds that swamp it.
This is costly to perform for all of our simulations, so instead we will define the difference 
\be
d_{1,2} (z) = \overline{T_{21}}^{(1)}(z) - \overline{T_{21}}^{(2)}(z)
\ee
between two GS models ($T_{21}^{(1)}$ and $T_{21}^{(2)}$, respectively), and simply compute the $\chi^2$ statistic
\be
\chi^2 = \sum_{i,j} d_{1,2}(z_i)  C^{-1}_{ij} d_{1,2}(z_j),
\label{eq:chisq}
\ee
as a metric of how different these two models are {\it in theory}.
Here the indices $i,j$ run over redshifts (or frequencies), and $C$ is the covariance matrix, where for our first analysis we can neglect the cosmic-variance component of $C$~\cite{Munoz:2020itp}, and take $C_{ij} = \sigma_i^2 \delta_{ij}$,
with an instrumental noise
\be
\sigma_i = \dfrac{T_{\rm sky}}{\sqrt{B t_{\rm obs}}},
\ee
determined by the observation time $t_{\rm obs}=1$ year, bandwidth $B=0.4$ MHz, 
and a sky temperature  $T_{\rm sky} (\nu) = 1570 \times (\nu/\nu_0)^{-2.5}$, anchored at $\nu_0=72$ MHz,
all chosen to closely match EDGES~\cite{Bowman:2018yin}.
Moreover, in this analysis we will consider the frequency range $\nu=50-110$ MHz, covering $z=12-27$, which covers the entire cosmic-dawn range of interest, and cuts off the beginning of reionization.

Before showing our results, let us emphasize that the $\chi^2$ obtained with Eq.~\eqref{eq:chisq} should be interpreted with caution.
This is for two main reasons.
First, we are not including any foreground marginalization, which can make two models appear more similar to each other, as well as diminish the overall significance of a prospective detection.
Second, we are keeping  all astrophysical parameters fixed, as varying them would dramatically increase the dimensionality of the problem, making it prohibitively expensive.
We will vary only one parameter, $\alpha$, which encapsulates the effect of feedback during cosmic dawn.
As a consequence, our reported $\chi^2$ values in this section ought to be interpreted as a theoretical best-case scenario of the difference between models, aimed to guide future detailed studies, whereas the specific values of $\chi^2$ will dampen when other effects are included.

\begin{figure}[t!]
	\includegraphics[width=0.5\textwidth]{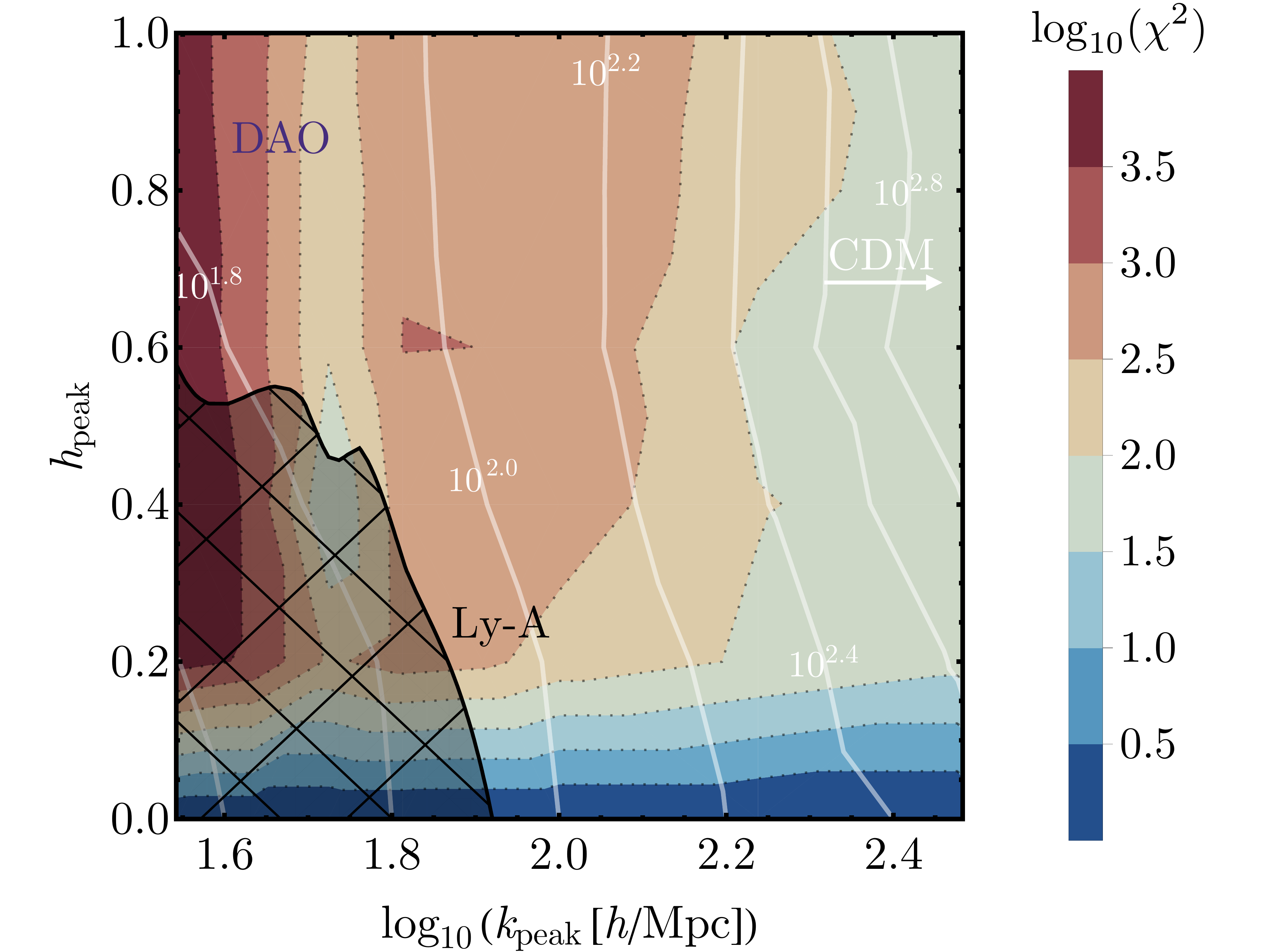}
	\caption{
	We show the comparison between each ETHOS model (as a function of its two effective parameters $k_{\rm peak}$ and $h_{\rm peak}$) and the closest WDM model (with different $k_{\rm peak}$ but $h_{\rm peak}=0$), using the 21-cm global signal.
	The color scale indicates the value of the $\chi^2$ difference between the two cases (which ought to be interpreted with care, see Eq.~\eqref{eq:chisq} and the surrounding discussion), whereas the white lines denote the	free-streaming scale for the closest WDM model (in units of $\log_{10}[k_{\rm peak}/(h$/Mpc)], see Eq.~\eqref{eq:kpeakWDM} for a translation to a WDM mass), which grows with $h_{\rm peak}$, as expected.
	The black shaded region is ruled out by Lyman-$\alpha$ data~\cite{Bohr:2020yoe,Murgia:2017lwo,Archidiacono:2019wdp}.
	}	
	\label{fig:2D_GS_WDM}
\end{figure}

We start by studying the differences between ETHOS and WDM models in the 21-cm GS.
In order to perform a meaningful comparison we will find the closest WDM model (with $h_{\rm peak}=0$ but $k_{\rm peak}<\infty$) to each ETHOS one, and report the $\chi^2$ difference between them.
For this, we interpolate the GS from our finite sample of WDM simulations to obtain results for arbitrary values of $k_{\rm peak}$.
We show the result of this analysis in Fig.~\ref{fig:2D_GS_WDM}.
As expected, low values of $h_{\rm peak}$ are very similar to WDM, and in fact for $h_{\rm peak}\leq 0.2$ the difference between WDM and ETHOS is small.
This difference grows for stronger DAOs, showing that the 21-cm signal has the potential to distinguish them from WDM.
Note that, at fixed $k_{\rm peak}$, higher values of $h_{\rm peak}$ produce less suppression, and thus the closest WDM model has a larger free-streaming scale.

\begin{figure}[btp!]
	\includegraphics[width=0.5\textwidth]{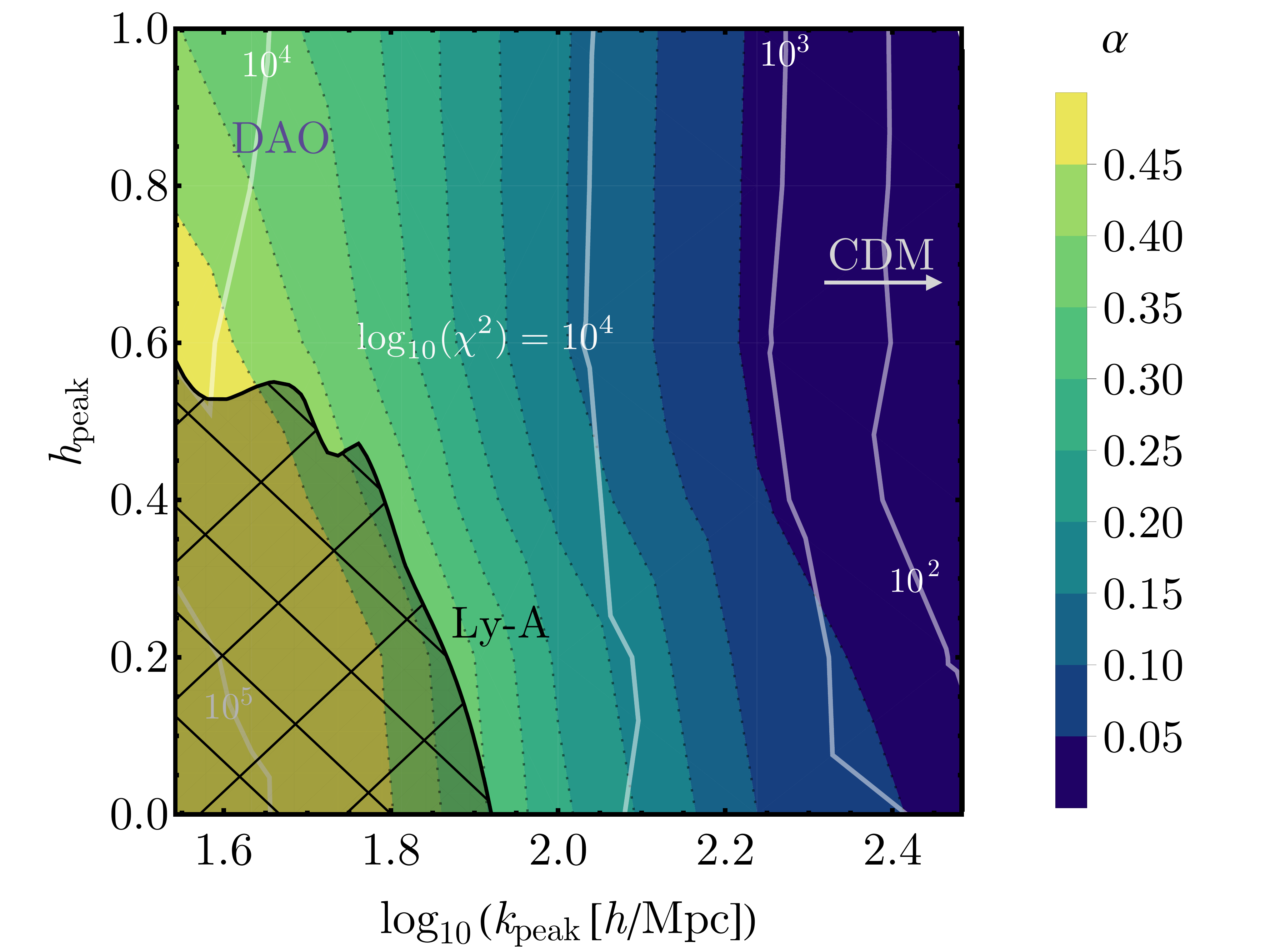}
	\caption{
	Same as Fig.~\ref{fig:2D_GS_WDM} but for the closest CDM+feedback model, parametrized by $\alpha$.
	Here the white lines indicate the $\chi^2$ difference, whereas the color scale follows the $\alpha$ of the closest CDM+feedback model. 
    The difference between CDM+feedback and ETHOS models grows with lower $k_{\rm peak}$. 
	}	
	\label{fig:2D_GS_alpha}
\end{figure}

We now move to find how different each ETHOS model is from CDM with feedback.
The analysis is similar to the WDM case, although now we interpolate between different values of  $\alpha=0-0.5$, which parametrizes the feedback strength.
We report the value of $\alpha$ that makes CDM closest to each ETHOS simulation, as well as the $\chi^2$ difference between them.
The results of this analysis are summarized in Fig.~\ref{fig:2D_GS_alpha}.
Larger values of $k_{\rm peak}$ in ETHOS correspond to more CDM-like behavior, and thus lower $\alpha$.
Interestingly, at fixed $k_{\rm peak}$ increasing the height $h_{\rm peak}$ of the DAOs requires lower $\alpha$, as there is more structure formation (and thus it is more similar to CDM).
The value of $\chi^2$ between the two models grows for smaller $k_{\rm peak}$, as warmer DM produces a more marked---and rapid---suppression than the smooth feedback.
Note that for $k_{\rm peak}\lesssim 10^{1.8}\,h/$Mpc the closest value of $\alpha$ saturates at 0.5, the maximum value we allow.

While in the comparison between ETHOS and WDM models (Fig.~\ref{fig:2D_GS_WDM}) the $\chi^2$ difference reached small values in part of the parameter space ($\lesssim 10$ for $h_{\rm peak}\leq0.2$), that is not the case when contrasting ETHOS and CDM+feedback.
Even for large values of $k_{\rm peak}$ we find a significant ($\chi^2\gtrsim 100$) deviation between  ETHOS and the closest CDM+feedback model.
These $\chi^2$ values would be reduced once foreground and astrophysical-parameter marginalization are included, as argued above.
Nevertheless, we expect that the relative size of these differences to hold, and thus that the ETHOS models that we explore are fairly  distinguishable from CDM+feedback, due to the cutoff nature of ETHOS suppression versus the smooth power-law suppression of the feedback, given the phenomenological feedback model that we have implemented.

\section{Effect on the 21-cm Fluctuations}
\label{sec:21cmPS}

In addition to the 21-cm GS, changing the HMF has a profound impact of the 21-cm fluctuations, which are expected to be measured soon by 21-cm interferometers~\cite{vanHaarlem:2013dsa,Eastwood:2019rwh,Beardsley:2016njr,Koopmans:2015sua, DeBoer:2016tnn}.
Let us now turn our attention to them.

\subsection{The observable}

We begin describing the 21-cm fluctuations and how we calculate them.
We use the same {\tt 21cmvFAST} simulation boxes from above, where we decompose the 21-cm temperature at each point as
\be
T_{21}(\mathbf x) = \overline{T_{21}} + \delta T_{21}(\mathbf x),
\ee
and calculate the Fourier-space two-point function of the 21-cm fluctuation $\delta T_{21}$.
This two-point function is the 21-cm power spectrum $P_{21}$.
For convenience we will employ the amplitude of 21-cm fluctuations, defined as
\be
\Delta^2_{21}(k_{21}) = \dfrac{k_{21}^3}{2\pi^2} P_{21}(k_{21}),
\ee
and refer to it as the 21-cm power spectrum (PS) unless confusion can arise.
In order to notationally differentiate the wavenumbers of 21-cm fluctuations from those of matter fluctuations, we refer to the former as $k_{21}$.
Interferometers such as the hydrogen epoch-of-reionization array (HERA) will probe the range $k_{21}\sim 0.1-1\,h/$Mpc,
as for lower wavenumbers foregrounds dominate, whereas for higher ones thermal noise does~\cite{DeBoer:2016tnn}.

\begin{figure}[t!]
	\includegraphics[width=0.5\textwidth]{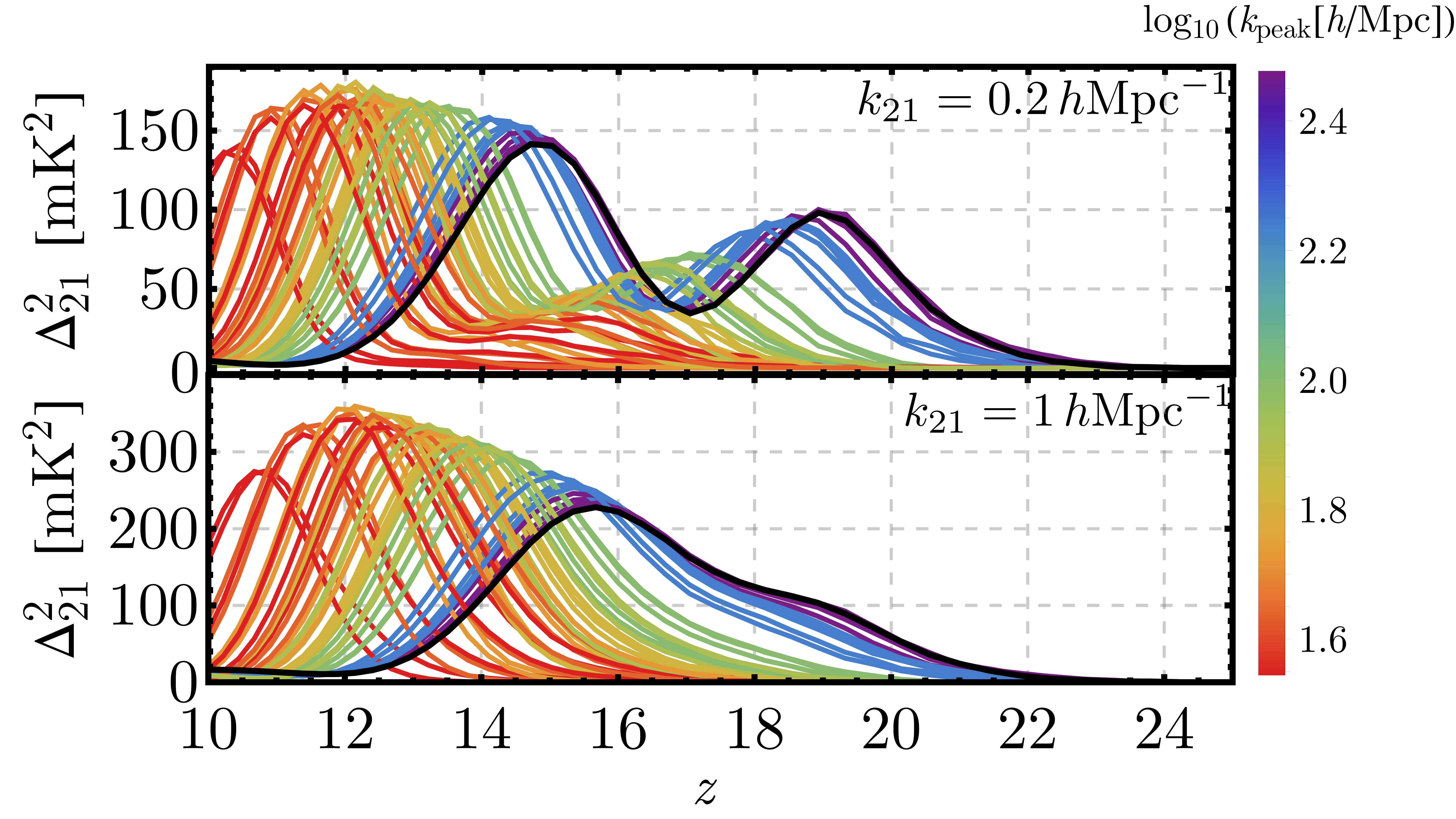}
	\caption{Amplitude of the 21-cm fluctuations as a function of redshift at two wavenumbers $k_{21}=0.2 h\,\Mpcinv$ ({\it top)} and $1 h\,\Mpcinv$ ({\it bottom)}.
	As in previous figures, the color encodes the wavenumber of the first peak $k_{\rm peak}$.
	}	
	\label{fig:PS_allETHOS}
\end{figure}

To build intuition, we show in Fig.~\ref{fig:PS_allETHOS} the 21-cm PS at two wavenumbers, $k_{21}=0.2$ and 1 $h/$Mpc, for all our ETHOS simulations.
These wavenumbers are chosen to represent both large- and small-scale 21-cm fluctuations that are observable by the current generation of experiments.
The origin of 21-cm fluctuations is different during each of the eras described above, so let us begin by describing the overall features of these curves.

We begin at early times, during the LCE ($z\sim 17-22$ for CDM), where fluctuations grow at all scales $k_{21}$.
That is because the UV photons are emitted from anisotropically distributed galaxies, and as they produce more WF coupling these fluctuations grow.
The large-scale (small $k_{21}$, top panel of Fig.~\ref{fig:PS_allETHOS}) fluctuations decrease in size during the transition between the LCE and the EoH ($z\sim17$ for CDM), as the effect of X-ray and UV photons cancel out~\cite{Munoz:2019rhi}, whereas at small scales (large $k_{21}$, bottom panel of Fig.~\ref{fig:PS_allETHOS}) there is no such cancellation.
Finally, the 21-cm fluctuations increase again during the EoH, until they nearly vanish by the time the gas is fully heated ($z\sim 12$ for CDM).
There will be a third peak at lower redshifts, due to reionization, which we do not consider, as we do not include lower redshifts in our analyses.

\begin{figure}[hbtp!]
	\includegraphics[width=0.5\textwidth]{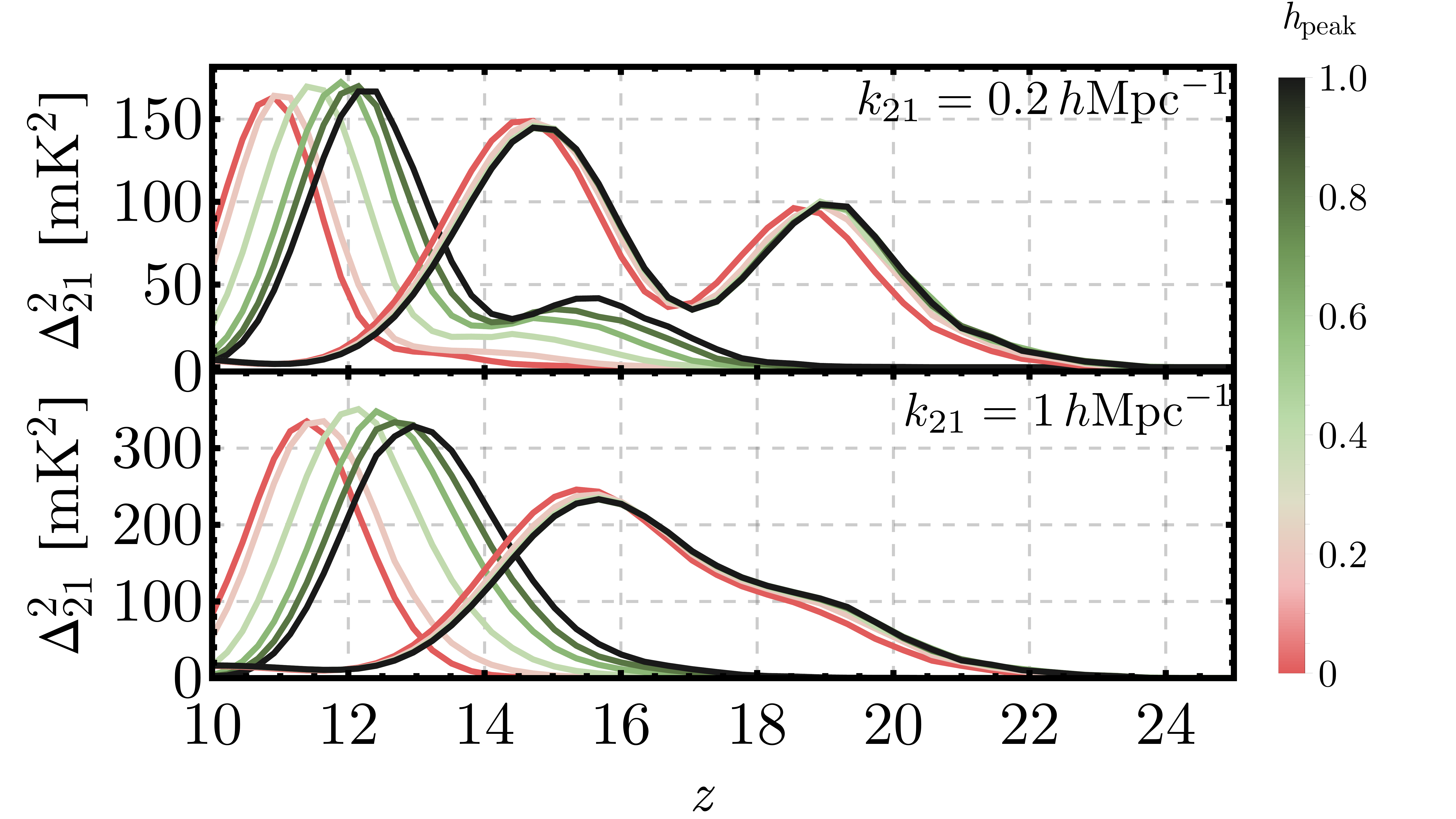}
	\caption{Same as Fig.~\ref{fig:PS_allETHOS}, but for the subset of models with $k_{\rm peak}= 43$ (left lines) and $300 \, h/$Mpc (right lines), with $h_{\rm peak}$ denoted by the line color.
	}	
	\label{fig:PS_hpeak}
\end{figure}

As was the case for the GS, ETHOS models show delayed structure formation, and thus the 21-cm PS curves shift to lower redshifts.
Nevertheless, the 21-cm fluctuations provide us with angular information, in the form of different $k_{21}$, which will allow us to better differentiate between models.
This is apparent, for instance, in Fig.~\ref{fig:PS_hpeak}.
There we show the 21-cm PS for models with two $k_{\rm peak}$, as a function of $h_{\rm peak}$.
The $h_{\rm peak}=0$ cases tend to form structure later than their higher-$h_{\rm peak}$ counterparts, as argued above.
Nevertheless, the shift in the high- and low-$k_{21}$ fluctuations is different.
For instance, the WDM ($h_{\rm peak}=0$) and full-DAO ($h_{\rm peak}=1$) curves with $k_{\rm peak}=43\,h/$Mpc in the top panel of Fig.~\ref{fig:PS_hpeak} have very different shapes, showing that the effect of DAOs is not just a shift, and the entire cosmic history of the 21-cm line can be used to differentiate between models.

\begin{figure}[t!]
	\includegraphics[width=0.5\textwidth]{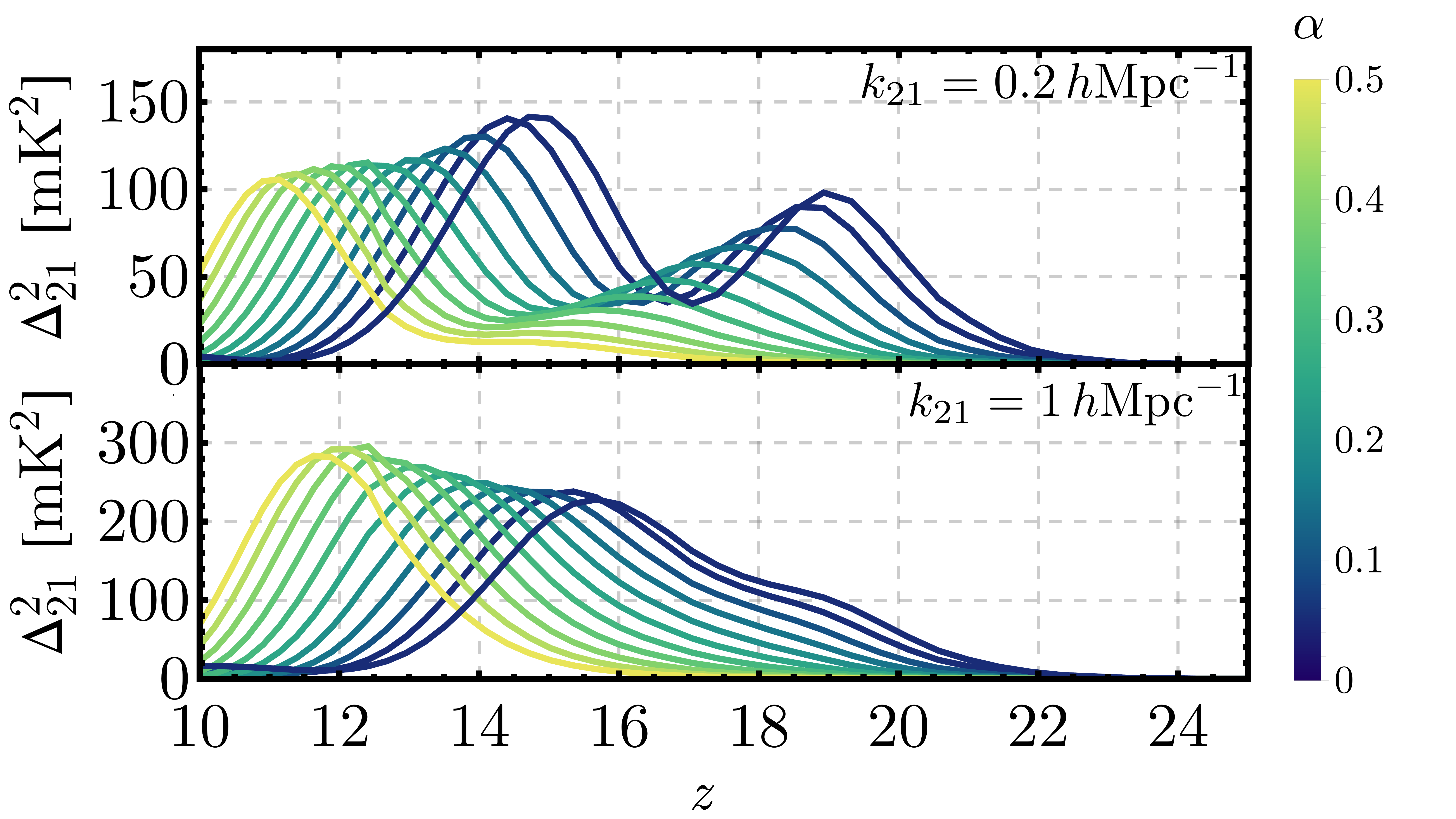}
	\caption{Same as Fig.~\ref{fig:PS_allETHOS}, but for CDM with varying amounts of feedback, parametrized through $\alpha$.
	}	
	\label{fig:PS_alpha}
\end{figure}

Finally, as we did before, we include CDM with feedback by varying the parameter $\alpha$ in Eq.~\eqref{eq:fstaralpha}.
We show the resulting power spectra in Fig.~\ref{fig:PS_alpha}, where as before larger $\alpha$ (stronger feedback) delay the onset of all the 21-cm transitions.
Interestingly, however, the 21-cm power is not just delayed, but its shape as a function of redshift also changes, owing to the impact that haloes of different masses have on the 21-cm line as a function of redshift~\cite{Munoz:2019hjh}.

\subsection{The Noise}

There are different ongoing and proposed 21-cm interferometers targeting the cosmic-dawn era~\cite{vanHaarlem:2013dsa,Eastwood:2019rwh,Beardsley:2016njr,Koopmans:2015sua, DeBoer:2016tnn}.
For concreteness, here we will focus on HERA~\cite{DeBoer:2016tnn}, and study how well it will be able to detect the fluctuations from all these models, as well as to distinguish them from one another and from CDM.
We will perform a realistic $\chi^2$ analysis here (as opposed to that in the previous section), using the noise expected of HERA.
We assume three years (540 days) of HERA data, and use the standard package {\tt 21cmSense}\footnote{\url{https://github.com/jpober/21cmSense}} to forecast the noise~\cite{Pober:2013jna,Pober2013}.
We discard all wavenumbers within the foreground wedge~\cite{Liu:2009qga,Morales:2012kf,Datta:2010pk,Parsons:2012qh}, whose extent we vary from an optimistic case, where the horizon limit is given by the experiment resolution, to a moderate and a pessimistic case, which include a supra-horizon buffer, following Ref.~\cite{Munoz:2019fkt} (see Appendix~\ref{app:Foregrounds} for more details).

A subtlety that we have to address is that, while the telescope (thermal) noise is the same for all of our simulations, they each have a different cosmic-variance (CV) noise, given their different fiducial power spectra.
This CV is important for low wavenumbers ($k_{21}\sim 0.1\,h\Mpcinv$), where thermal noise is small.
Instead of running {\tt 21cmSense} for each of our  simulations individually, which is computationally slow, we devise a way of  including CV for any arbitrary 21-cm PS quickly but exactly.
The full noise of the 21-cm PS can be expressed as a sum of the thermal (th) and CV components, where the former is independent of the 21-cm model assumed, and the latter can be described as
$\sigma_{\rm CV}(\Delta^2_{21}) = a_{21} \times \Delta^2_{21}
$
for some $a_{21}$ that depends on $k_{21}$ and $z$, and varies with the experimental setup, but not with $\Delta^2_{21}$.
Thus, we calibrate this $a_{21}$ by using {\tt 21cmSense}, and find the full error as
\be
\sigma_{\rm full}(\Delta^2_{21}) = \sigma_{\rm th} + a_{21}\, \Delta^2_{21},
\ee
for each 21-cm PS $\Delta^2_{21}$,
where we have suppressed the dependence on $k_{21}$ and $z$ of all terms in that equation.
We have confirmed that this expression exactly recovers the full noise when using different input 21-cm power spectra in {\tt 21cmSense}.

\begin{figure}[hbtp!]
	\includegraphics[width=0.48\textwidth]{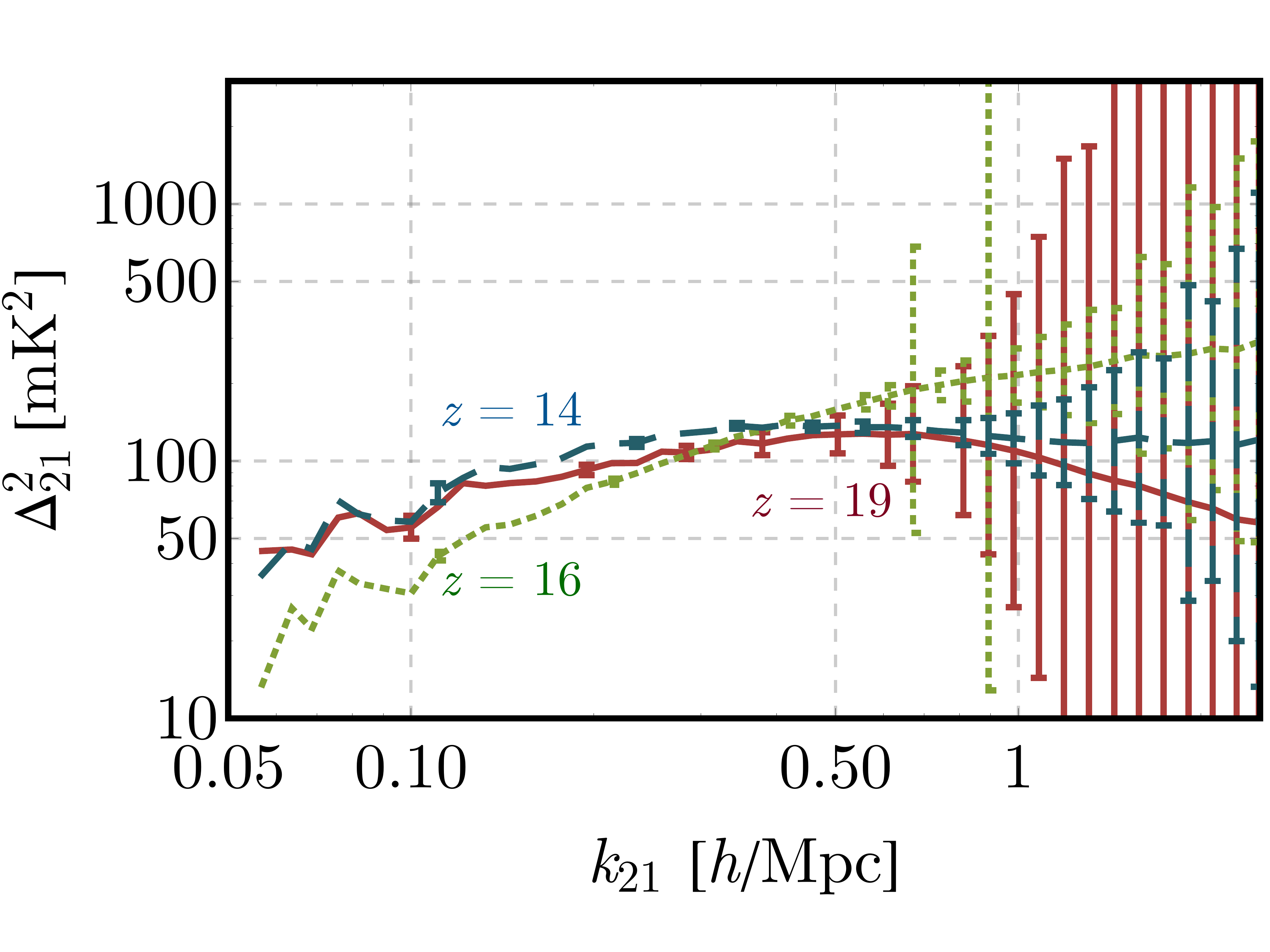}
	\caption{Amplitude of 21-cm fluctuations as a function of wavenumber for our CDM model (with $\alpha=0$), as well as the forecasted noise for 540 days of HERA data, assuming moderate foregrounds.
	We show the results at three redshifts, roughly corresponding to the peak of the LCE ($z=19$), the transition to the EoH ($z=16$), and the peak of the EoH ($z=14$) for this model.
	Wavenumbers without an errorbar cannot be measured at any precision.
	}	
	\label{fig:P21_k21_noise}
\end{figure}

In order to perform our analysis we divide the frequency range $\nu=50-120$ MHz in bins that are 4 MHz in size.
These are wider than for the GS, as we ought to average over more cosmic volume to bring the noise per $k_{21}$ mode down at each $z$.
We show the expected noise for our CDM model, under moderate foregrounds, in Fig.~\ref{fig:P21_k21_noise}.
We will analyze wavenumbers in the range $k_{21}=0.05-2.5\,h/$Mpc, though the majority of modes do not have a measurement, as clear in Fig.~\ref{fig:P21_k21_noise}, due to the foreground wedge. 
For low $k_{21}$ only a handful of modes can be observed, although they can reach small errors as they are observed many times.
For larger $k_{21}$ (smaller scales), however, the situation is reversed, and more modes with $k_{21}\gtrsim 0.5 \,h/$Mpc can be observed, while they each have large noise.

\begin{figure}[btp!]
	\includegraphics[width=0.48\textwidth]{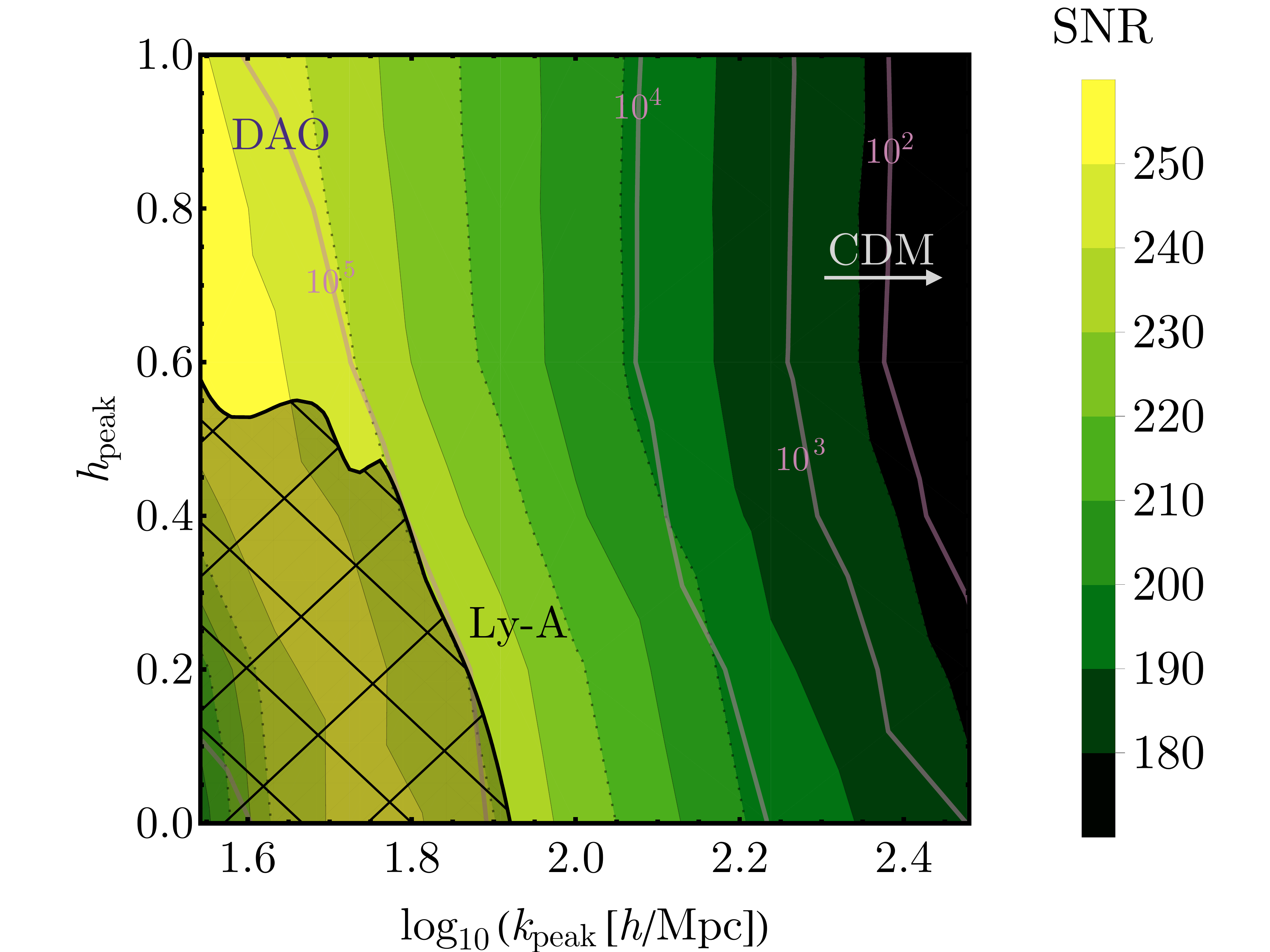}
	\caption{
	Forecasted signal-to-noise ratio (SNR) of the 21-cm PS for different ETHOS models, in color map. 
	In all cases we assume 540 days of HERA data, and moderate foregrounds.
	The thin purple lines follow the contours of constant $\chi^2$ difference between each ETHOS model and CDM (with no feedback), which grows towards the left of the plot.
	}	
	\label{fig:2Dplot_SNR_PS}
\end{figure}

\subsection{Detectability}

We will use two metrics to study how detectable---and differentiable from each other---our ETHOS models are.
The first is the signal-to-noise ratio (SNR), and the second is the $\chi^2$ statistic.
In all cases we will assume a diagonal covariance matrix, ignoring correlations between different $k_{21}$ and $z$ bins, for simplicity.

We begin by calculating the SNR for each of our models, computed through
\be
{\rm SNR}^2 = \sum_{i_k,i_z} \dfrac{\Delta^2_{21}(k_{21},z)}{\sigma^2_{\rm full}(k_{21},z)},
\ee
where the sum runs over all wavenumber $i_k$ and redshift $i_z$ bins.
We show the SNR for all our models, assuming moderate foregrounds, in Fig.~\ref{fig:2Dplot_SNR_PS}.
We find SNR $\approx 150-250$, varying smoothly as a function of the ETHOS parameters.
Interestingly, more-suppressed ETHOS models haver higher SNR than their CDM-like counterparts.
The reason is that a stronger suppression of power delays structure formation, and moves all the 21-cm landmarks to lower $z$, where the noise is smaller (as $T_{\rm sky}$ sharply rises at lower frequencies---or high redshifts).
This trend is reversed for ETHOS models with $k_{\rm peak}\lesssim10^{1.6} \,h/$Mpc, however, as their cosmic-dawn evolution is late enough that it is not completed by $z=10$, when our simulations end.
Nevertheless, the models for which this is true are already in tension with Lyman-$\alpha$ observations~\cite{Bohr:2020yoe}, as clear in Fig.~\ref{fig:2Dplot_SNR_PS}.

As all our ETHOS models are detectable at high SNR, we now perform a $\chi^2$ test to distinguish between them, similar to the previous section.
Given the difference $\Delta^2_{\rm diff}$  between the 21-cm power spectra of two models, we define their $\chi^2$ to be 
\be
\chi^2 = \sum_{i_k,i_z} \dfrac{\Delta^2_{\rm diff}(k_{21},z)}{\sigma^2_{\rm full}(k_{21},z)},
\ee
where the noise in the denominator is evaluated for the first of the two models (which will always be the one plotted).
While this $\chi^2$ for the 21-cm PS shares some of the same caveats as that of the GS (as we are not simultaneously varying astrophysical parameters due to the computational cost), it is fundamentally more robust.
The reason for that is twofold.
First, here we do not have to subtract foregrounds, as we only consider data outside of the wedge, which is expected to be foreground clean.
Second, here we are taking realistic forecasted noises for HERA, as opposed to using the ``ideal" radiometer equation for the GS, which results in lower overall values of the $\chi^2$ for the PS than for the GS, thought these can be trusted more.

Looking at Fig.~\ref{fig:2Dplot_SNR_PS} once more, we see that essentially all ETHOS models are very different from the vanilla CDM scenario, as the $\chi^2$ difference between them is always larger than 10, and grows dramatically as $k_{\rm peak}$ decreases, especially below $10^{2.4}\,h/$Mpc.
However, as argued above, some of this difference can be absorbed by a difference in the astrophysics.
Moreover, we want to know if ETHOS models can be distinguished from WDM given a fiducial 21-cm observation. 
We now tackle these two questions.

\begin{figure}[hbtp!]
	\includegraphics[width=0.48\textwidth]{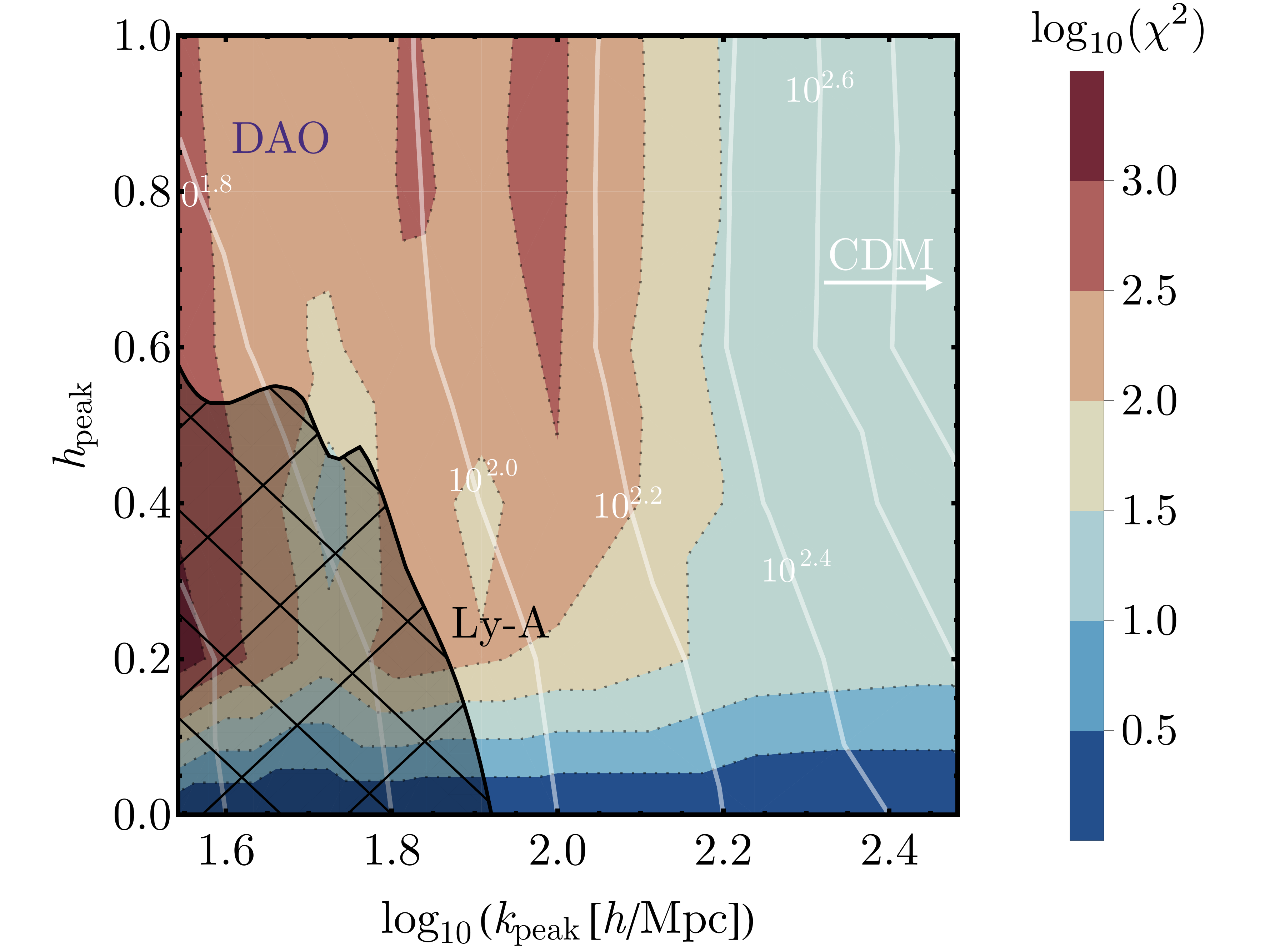}
	\caption{
	Different between each ETHOS model and the closest WDM case, as in Fig.~\ref{fig:2D_GS_WDM}, but for the 21-cm fluctuations, assuming 540 days of HERA data and moderate foregrounds.
	The $\chi^2$ differences reported here (as white contours) are more robust to marginalization than those in Fig.~\ref{fig:2D_GS_WDM}.
	}	
	\label{fig:2Dplot_bfWDM_PS}
\end{figure}

We begin, as in the previous section, by comparing ETHOS models with DAOs against their closest WDM counterpart.
We show the summary of this analysis in Fig.~\ref{fig:2Dplot_bfWDM_PS}.
As before, we find that at fixed $k_{\rm peak}$ models with strong DAOs (large $h_{\rm peak}$) suppress structure less.
Now, however, the $\chi^2$ difference between models is slightly smaller, and in fact it is below 10 for $h_{\rm peak}<0.2$, making those barely distinguishable from WDM.
Moreover, all models with $k_{\rm peak}>10^{2.2}\,h/$Mpc have differences $\chi^2 \lesssim 30$ with respect to their closest WDM counterpart, as such small scales chiefly affect high redshifts where the PS noise is too high to distinguish them.
On the opposite side, the difference between models grows for larger values of $h_{\rm peak}$, especially at low $k_{\rm peak}$.
For instance the larger-scale DAOs, with $h_{\rm peak}\gtrsim 0.4$ and $k_{\rm peak}\lesssim 100\,h/$Mpc, give rise to large $\chi^2\sim\mathcal O(100)$ differences, and thus could be promptly distinguished from WDM.
This shows the promise of 21-cm PS measurements to detect and characterize DAOs.

\begin{figure}[hbtp!]
	\includegraphics[width=0.48\textwidth]{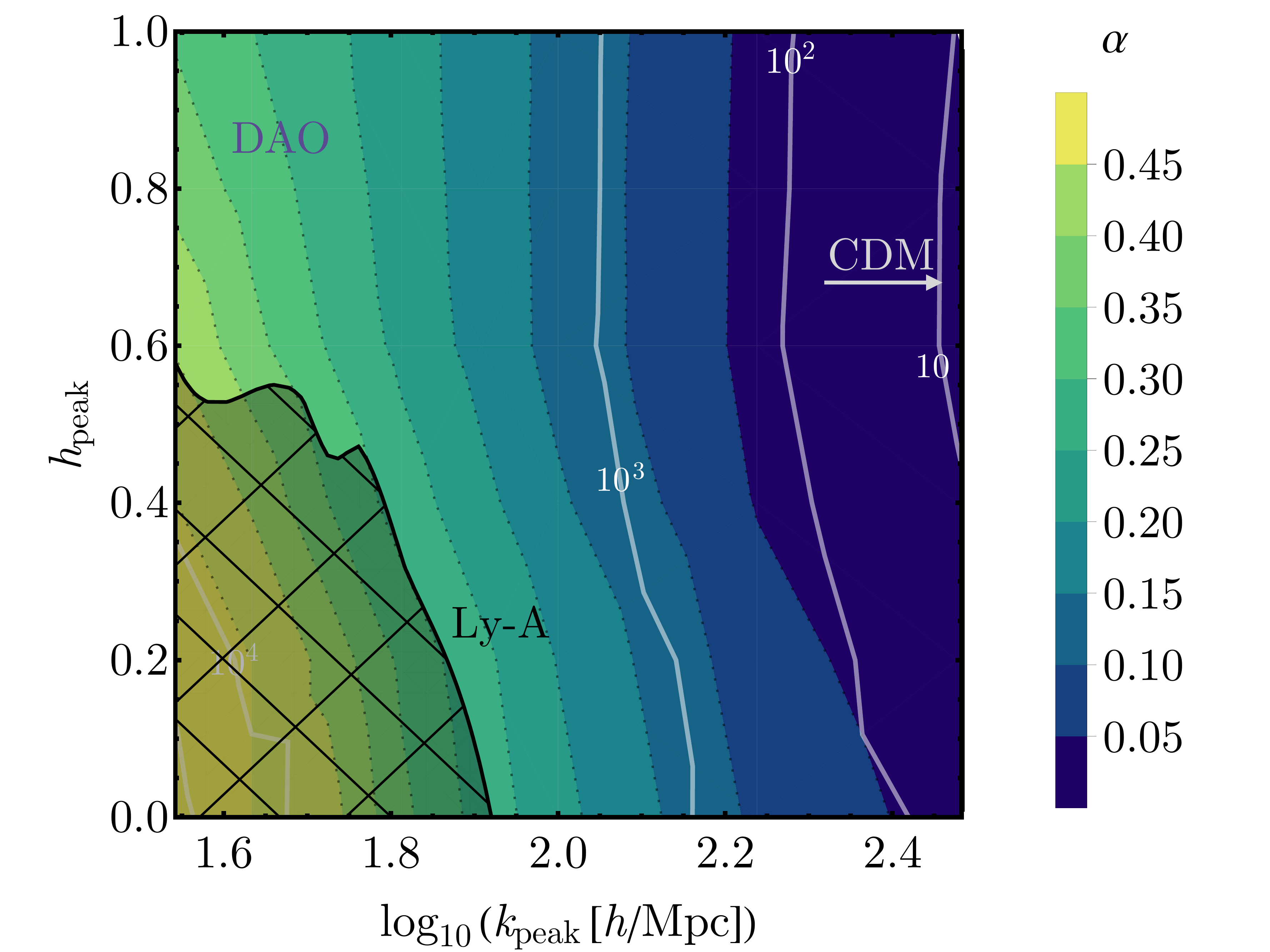}
	\caption{
	Same as Fig.~\ref{fig:2Dplot_bfWDM_PS} but comparing each ETHOS model to the closest CDM including feedback, whose strength is parametrized through $\alpha$.
	}	
	\label{fig:2Dplot_bfalpha_PS}
\end{figure}

Additionally, we study how well HERA could distinguish ETHOS models from CDM+feedback.
A summary of our findings is in Fig.~\ref{fig:2Dplot_bfalpha_PS}.
As before, ETHOS models with more suppression (lower $k_{\rm peak}$) are matched to CDM models with stronger feedback (larger $\alpha$).
However, here the low-$k_{\rm peak}$ range can be better distinguished from CDM+feedback than when using the GS, given the additional information from different wavenumbers.
For the same reason, the best-fit values of the feedback-strength $\alpha$ for each ETHOS model are slightly different for the 21-cm PS than for the GS.
As was the case in Fig.~\ref{fig:2Dplot_bfWDM_PS}, the high-$k_{\rm peak}$ part of the parameter space is more difficult to probe with the 21-cm PS, as those models show their most marked suppression at high redshifts, where the noise is large.
Nevertheless, we find that ETHOS models with $k_{\rm peak}\lesssim 10^{2.3}\,h/$Mpc give rise to a $\chi^2$ difference larger than 100,
showing that HERA has the potential to tell ETHOS apart from CDM+feedback, given our model assumptions.

Throughout this section we have shown results assuming moderate foregrounds, where the vast majority of 21-cm modes observed by HERA would be within the foreground wedge, and thus unusable for our analysis.
The extent of the wedge is, as of yet, uncertain at the redshifts we consider, so we have re-done our analyses assuming two other different foreground options, an optimistic one and a pessimistic one.
We show the results in Appendix~\ref{app:Foregrounds}, and simply summarize them here.
We find that pessimistic foregrounds reduce the SNR of a prospective 21-cm PS detection by  roughly $10\%$ for all ETHOS models, as well as CDM, whereas the optimistic-foreground assumption increases the SNR by roughly a factor of 2.
We additionally find that the $\chi^2$ comparisons follow a similar trend as in the moderate-foreground case considered in the main text, though a factor of $\sim 5$ worse (better) for pessimistic (optimistic) foregrounds.
This would change the specific cut of the ETHOS parameter space that is distinguishable from CDM+feedback or WDM, but would not alter our main conclusions.

\section{Discussion and Conclusions}
\label{sec:Conclusions}

In this work we have carried out an exploratory study of how upcoming measurements of the 21-cm line of hydrogen during cosmic dawn can determine the nature of the dark sector, through the small-scale behavior of DM.
For that, we have followed the ETHOS paradigm, which translates the microphysical degrees of freedom of the DM and DR interactions into two key variables: the location $k_{\rm peak}$ and amplitude $h_{\rm peak}$ of the first DAO peak.
We carried out $N$-body simulations of each ETHOS model to find their halo mass function down to the atomic-cooling threshold, and used those as input of semi-numeric {\tt 21cmvFAST} simulations to find the evolution of the 21-cm signal from the formation of the first stars to reionization.
We then studied the prospects to detect, and distinguish, ETHOS models with upcoming measurements of the 21-cm global signal and fluctuations.

Our results can be summarized as follows.
All ETHOS models with a suppression scale $k_{\rm peak}\lesssim 10^{2.5}\,h/$Mpc can be distinguished from CDM by both the 21-cm GS and the PS, even when varying the strength of the feedback processes in CDM.
More interestingly, in the case that a prospective 21-cm detection shows a lack of power at high $k$, we have shown that ETHOS models with $h_{\rm peak}\gtrsim 0.4$ can be differentiated from WDM.
That is because the cutoff in WDM produces a more sudden turn-on of the 21-cm signal than ETHOS models with strong DAOs, which exhibit a bump in power at smaller scales.
Moreover, even models with DAOs can be distinguished from our feedback model, as this is expected to only suppress stellar formation in a smooth manner, rather than the sharper cut of non-CDM models.

Ours is the first study of the evolution of the 21-cm signal across cosmic dawn including DAOs of different heights and locations. 
As such, we have taken some simplifying assumptions to timely explore the large ETHOS parameter space.
First, we have not considered small-mass molecular-cooling haloes, as resolving those requires finer-resolution $N$-body simulations.
Nevertheless, as those haloes are formed out of smaller-scale fluctuations deviations from the standard CDM paradigm will be more apparent, and our analysis is, therefore, conservative.
Second, we have only varied one astrophysical parameter (the strength of the stellar feedback in CDM), instead of freely allowing all possible parameters in {\tt 21cmvFAST} to vary.
Last, in our global-signal forecasts we have ignored foreground marginalization.
These simplifying assumptions will be relaxed in subsequent work.
Throughout this paper we have assumed some fiducial observation time of 1000 hours for a global-signal experiment, and 4320 hours for a 21-cm fluctuation experiment.
These were chosen for convenience only, and our results can be trivially rescaled for different observation times $t_{\rm obs}$.
Despite these caveats, this work is a proof-of-concept that data of the 21-cm line of hydrogen at high redshifts ($z\approx10-25$) can readily distinguish different ETHOS models from the standard CDM, as well as from each other, probing a large swath of parameter space that is currently open.

In summary, we have shown that the cosmic-dawn era holds a trove of information about the small-scale behavior of matter fluctuations.
A detection of the 21-cm signal will, therefore, open the window to understanding the nature of DM in a regime currently unprobed, shedding light onto the nature of the dark sector.

\acknowledgements

It is our pleasure to thank 	Torsten Bringmann and Christoph Pfrommer for comments on a previous version of this draft.
JBM was supported by NSF grant AST-1813694
at Harvard and the Clay Fellowship at the Smithsonian
Astrophysical Observatory.
SB and JZ were funded by a Grant of Excellence from the Icelandic Research Fund (grant number 173929).
MV acknowledges support through NASA ATP grants 16-ATP16-0167, 19-ATP19-0019, 19-ATP19-0020, 19-ATP19-0167, and NSF grants AST-1814053, AST-1814259,  AST-1909831 and AST-2007355.
All simulations were run on the FASRC Cannon cluster supported by the FAS Division of Science Research Computing Group at Harvard University.

\bibliography{ETHOS21cm,dark_matter_ref}

\appendix

\begin{figure}[btp!]
	\includegraphics[width=0.48\textwidth]{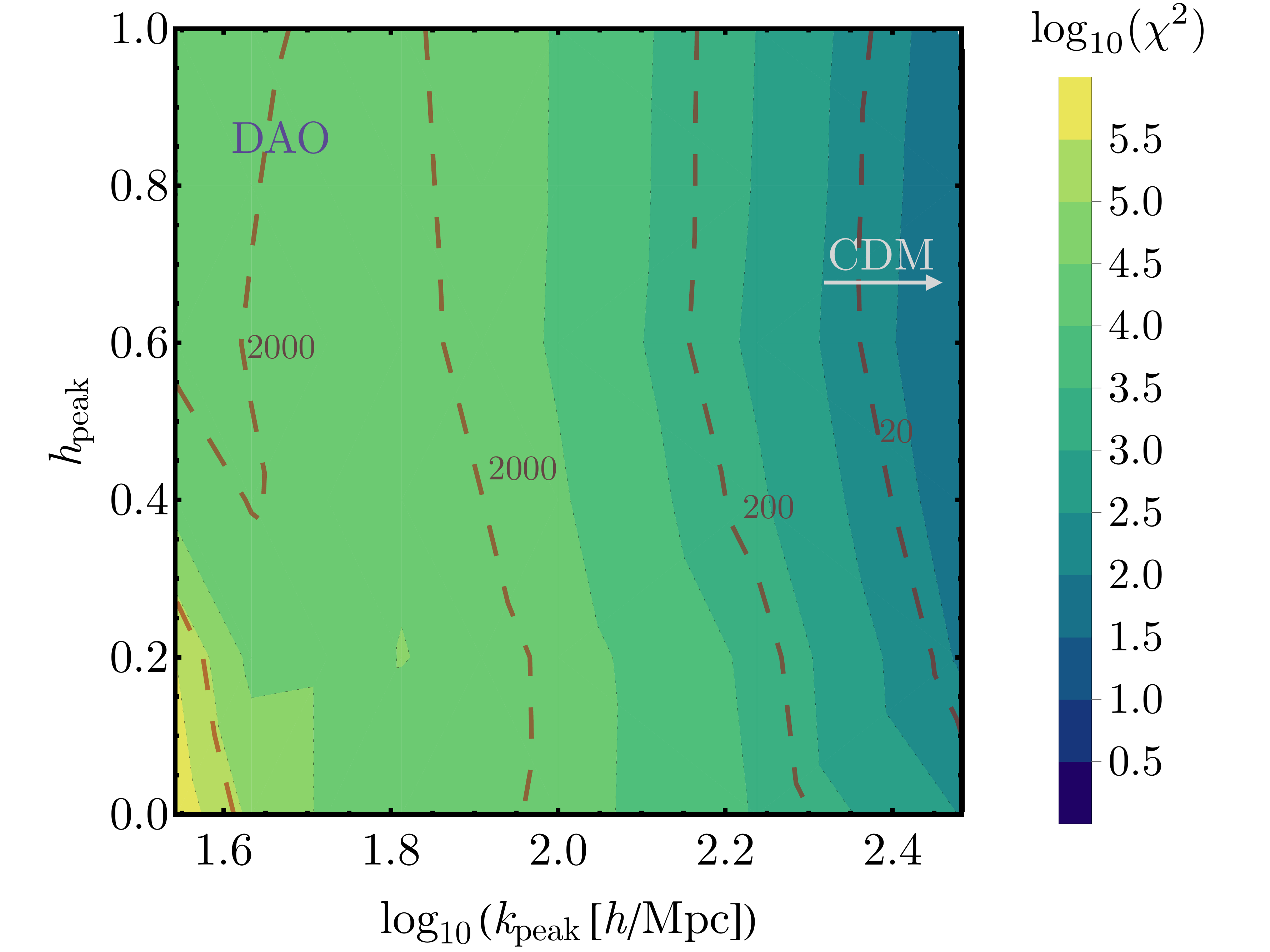}
	\caption{
	Difference in $\chi^2$ between each ETHOS model and the closest CDM+feedback, assuming pessimistic (dashed brown contours) and optimistic foregrounds (color map).  
	We do not show the best-fit $\alpha$ here, since they are visually very similar to Fig.~\ref{fig:2Dplot_bfalpha_PS}.
	}	
	\label{fig:2Dplot_bfalpha_PS_OPT_PESS}
\end{figure}

\section{Foregrounds in the 21-cm Power Spectrum}
\label{app:Foregrounds}

In this appendix we describe alternatives for the extent of the foreground wedge, which determines which wavenumbers can be measured by the 21-cm power spectrum, and to which precision.
We take a simple model of the foreground wedge, where wavenumbers along the line of sight ($k_{||}$) with 
\be
k_{||} \leq a(z) + b(z) k_\perp,
\label{eq:wedge}
\ee
are considered to be contaminated by foregrounds, and are thus unusable for our DM studies.
The two parameters $a$ and $b$ determine the extent of the wedge (see Refs.~\cite{Pober2013,Pober:2013jna} for details and its the implementation in {\tt 21cmSense}) as a function of the perpendicular wavenumber $k_\perp$, where $b(z)$ determines the extent of the horizon, and $a(z)$ accounts for a supra-horizon buffer where foregrounds may leak out~\cite{Orosz:2018avj}.
We take three three assumptions for the foreground wedge, following Ref.~\cite{Munoz:2019fkt}.
In the main text we assumed moderate foregrounds, which is our best guess for the extent of the wedge.
Here, instead, we explore what the results would be if foregrounds were more optimistic, where $b$ is given by the primary beam and we take no buffer ($a=0$), and a more pessimistic case where $a=0.1\,h/$ Mpc (instead of half of that in the moderate case).

\begin{figure}[t!]
	\includegraphics[width=0.48\textwidth]{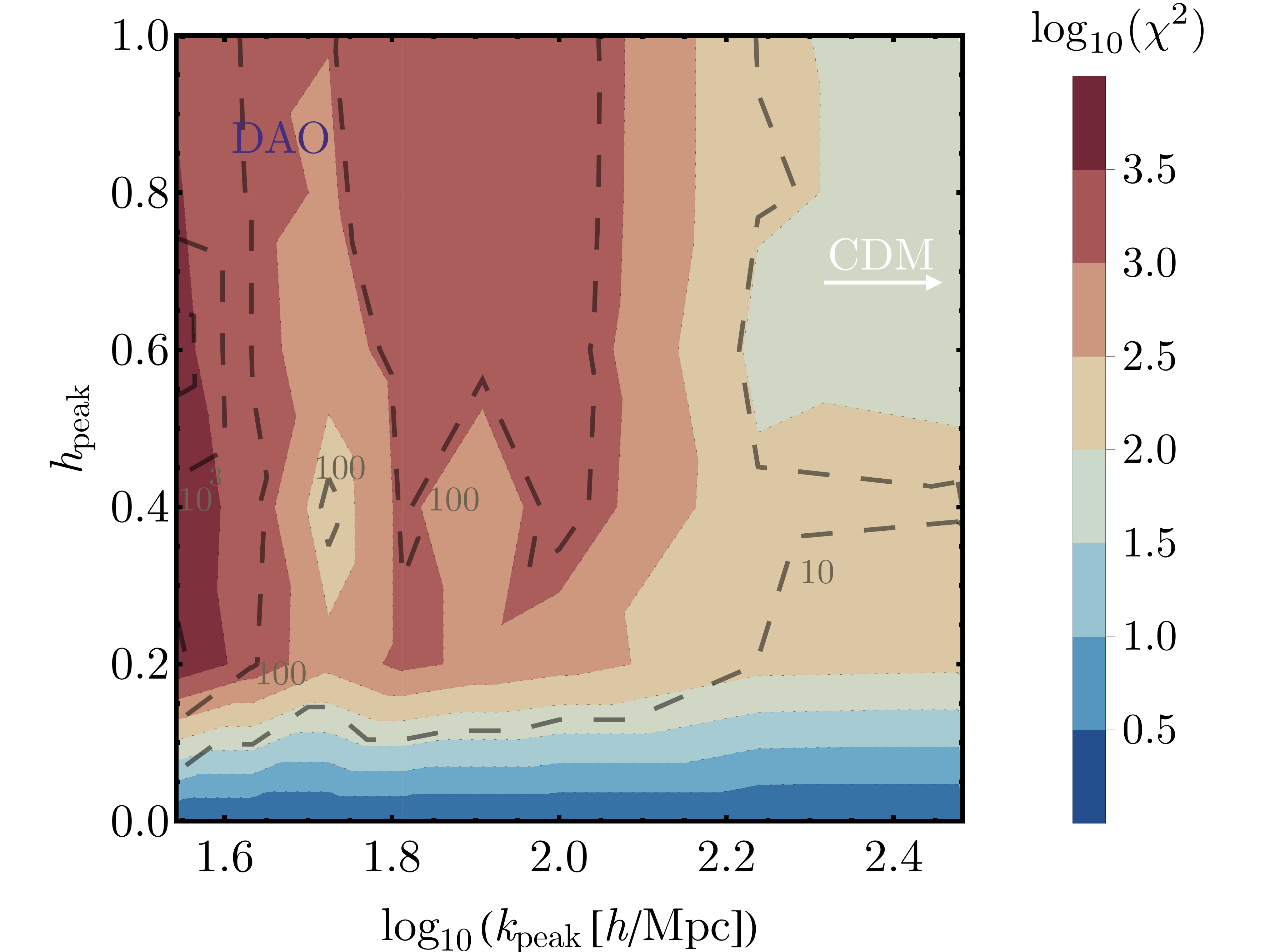}
	\caption{
	Same as Fig.~\ref{fig:2Dplot_bfalpha_PS_OPT_PESS} but comparing against the closest WDM case to each ETHOS model.
	}	
	\label{fig:2Dplot_bfWDM_PS_OPT_PESS}
\end{figure}

We show our results for these two foreground assumptions in Figs.~\ref{fig:2Dplot_bfalpha_PS_OPT_PESS} and~\ref{fig:2Dplot_bfWDM_PS_OPT_PESS}.
The first of these Figures shows the detectability of ETHOS models against CDM and feedback. We find that for the pessimistic-foregrounds case the expected $\chi^2$ is only a factor of $\sim 2$ worse than for the moderate case.
Assuming optimistic foregrounds, however, changes the picture significantly, as the large amount of wavenumbers $k_{21}$ accessible, and the great precision for each of them, allows all ETHOS models we study to be distinguishable from CDM and feedback at $\chi^2>100$.
The situation is similar in the comparison with WDM, shown in Fig.~\ref{fig:2Dplot_bfWDM_PS_OPT_PESS}.
Pessimistic foregrounds can still differentiate ETHOS models from WDM at $\chi^2>10$ for $h_{\rm peak}\geq0.4$, as long as $k_{\rm peak}\leq 10^2\,h/$Mpc.
Here, again, optimistic foregrounds would open a larger swath of parameter space, as only models with $h_{\rm peak}<0.1$ can be confounded with WDM in that case.
This shows that great progress can be made even when all 21-cm modes within the foreground wedge are discarded, yet the gains from recovering those modes would dramatically enhance our understanding of the dark sector.

\end{document}